\documentclass[twocolumn]{aastex63}
\usepackage[figuresright]{rotating}
\usepackage{subfigure}
\usepackage{epic,eepic}
\usepackage{graphicx}
\usepackage{threeparttable}
\shorttitle{ELM}
\shortauthors{Yuan,H.L., et al.}

\begin{document}

\title{ELM of ELM-WD: An extremely low mass hot star discovered in LAMOST survey}


\author{Hailong Yuan{$^{\dag}$}}
\affiliation{Key Laboratory of Optical Astronomy, National Astronomical Observatories, Chinese Academy of Sciences, Beijing 100101, China}

\author{Zhenwei Li}
\affiliation{Yunnan Observatories, Chinese Academy of Sciences, 650011, China}

\author{Zhongrui Bai}
\affiliation{Key Laboratory of Optical Astronomy, National Astronomical Observatories, Chinese Academy of Sciences, Beijing 100101, China}

\author{Yiqiao Dong}
\affiliation{Key Laboratory of Optical Astronomy, National Astronomical Observatories, Chinese Academy of Sciences, Beijing 100101, China}

\author{Mengxin Wang}
\affiliation{Key Laboratory of Optical Astronomy, National Astronomical Observatories, Chinese Academy of Sciences, Beijing 100101, China}

\author{Sicheng Yu}
\affiliation{Key Laboratory of Optical Astronomy, National Astronomical Observatories, Chinese Academy of Sciences, Beijing 100101, China}

\author{Xuefei Chen}
\affiliation{Yunnan Observatories, Chinese Academy of Sciences, 650011, China}

\author{Yongheng Zhao}
\affiliation{Key Laboratory of Optical Astronomy, National Astronomical Observatories, Chinese Academy of Sciences, Beijing 100101, China}

\author{Yaoquan Chu}
\affiliation{University of Science and Technology of China, Hefei 230026, China}

\author{Haotong Zhang{$^{\star}$}}
\affiliation{Key Laboratory of Optical Astronomy, National Astronomical Observatories, Chinese Academy of Sciences, Beijing 100101, China}

\email{{$\dagger$} yuanhl@bao.ac.cn}
\email{{$\star$} htzhang@bao.ac.cn}

\begin{abstract}

The Extremely Low Mass White Dwarfs (ELM WDs) and pre-ELM WDs are helium core white dwarfs with mass $<\sim 0.3M_{\odot}$.
Evolution simulations show that 
a lower mass limit for ELM WDs exists at $\approx0.14M_{\odot}$ and
no one is proposed by observation to be less massive than that.
Here we report the discovery of a binary system, 
LAMOST J224040.77-020732.8 (J2240 in short), 
which consists of a very low mass hot star and a compact companion.
Multi-epoch spectroscopy shows an orbital period $P_{orb} =$0.219658$\pm0.000002$ days 
and a radial velocity semi-amplitude $K1=318.5\pm3.3km/s$, 
which gives the mass function of 0.74$M_{\odot}$, indicating the companion is a compact star.
The F-type low resolution spectra illustrate no emission features, 
and the temperature ($\sim 7400K$) is consistent with that from Spectral Energy Distribution fitting and multi-color light curve solution.
The optical light curves,
in ZTF g, r and i bands and Catalina V band,
show ellipsoidal variability with amplitudes $\sim30\%$, 
suggesting that the visible component is heavily {\bf tidally} distorted.
Combining the distance from Gaia survey,
the ZTF light curves are modeled with Wilson-Devinney code and the result shows that
the mass of the visible component is $M1=0.085^{+0.036}_{-0.024}M_{\odot}$,
and the mass of the invisible component is $M2=0.98^{+0.16}_{-0.09}M_{\odot}$.
The radius of the visible component is $R1=0.29^{+0.04}_{-0.03}R_{\odot}$.
The inclination angle is approximately between 60$^{\circ}$ and 90$^{\circ}$.
The observations indicate the system is most likely a pre-ELM WD + WD/NS binary,
and the mass of pre-ELM is possibly lower than the $0.14M_{\odot}$ theoretical limit.


\end{abstract}


\section{INTRODUCTION}
\label{intro.sec}

ELM WDs or pre$-$ELM WDs are helium core white dwarfs 
with mass below 
0.30$M_{\odot}$ \citep[][]{2019ApJ...871..148L}.
Given their extreme low mass and the current age of our Universe, 
they can only be formed within 
interacting binary systems \citep{2010ApJ...723.1072B, 2018MNRAS.475.2480P}.
The progenitor of the ELM WD are thought
to be a low mass star with mass between 1.0$M_{\odot}$ and 1.5$M_{\odot}$, being stripped off its hydrogen envelope
by its companion either through the stable Roche lobe overflow channel (RL channel) or the common envelope ejection channel (CE channel) \citep[][]{2002MNRAS.336..449H, 2017MNRAS.467.1874C, 2018ApJ...858...14S, 2019ApJ...871..148L}.
ELM WDs with mass lower than 0.18$M_{\odot}$ are thought to be produced only in RL channel, since in the CE channel, the thick envelope can't be thrown off and the two star will merge when 
the mass of the helium core is lower than 0.21$M_{\odot}$ \citep[][]{2018ApJ...858...14S, 2019ApJ...871..148L}.
The pre-ELM phase happened at the end of the mass transfer of the RL channel, where the donor may still bear a thick 
bloated hydrogen envelope 
sustained by hydrogen burning for up to Gyrs before they contract to white dwarf cooling track \citep[][]{2014A&A...571A..45I}.
Those bloated pre-ELMs have similar color as main-sequence A and F star and hard to be distinguished by color or low resolution spectra \citep[][]{2017ApJ...839...23B,2017ASPC..509..447P,2018MNRAS.475.2480P,2021MNRAS.505.2051E}.
The mass of ELM WD formed in the RL channel lies between 0.14 and 0.4 $M_{\odot}$. The minimum mass is set by the bifurcation initial period \citep[][]{1988A&A...191...57P, 2019ApJ...871..148L}, below which the donor cannot develop a compact core and {\bf will likely evolve to a brown dwarf} \cite[][]{2004ApJ...616.1124N,2017MNRAS.467.1874C}. 
Different evolution calculations \citep[e.g.][]{2016A&A...595...35,2017MNRAS.467.1874C,2018ApJ...858...14S,2019ApJ...871..148L,2021MNRAS.506.3266S} show that the lower limit of ELM WD is around 0.14-0.16 $M_{\odot}$.
The observational mass limit seems to coincident with the theoretical result \citep[e.g.][]{2016ApJ...818..155B}. 
However, it is necessary to mention that the observational limit is not as well established as the theoretical one, given the difficulties separating pre-ELMs from main sequence stars.
\citet{2021MNRAS.508.4106E} listed the 21 spectral confirmed pre-ELM WD systems, 6 have masses lower than 0.14 $M_{\odot}$, but still consistent with the theoretic estimation considering their uncertainties.

In this work we report a very low mass pre-ELM WD, J2240, discovered in LAMOST low resolution spectral survey. With the help of LAMOST radial velocity {\bf (RV)} curve, GAIA Dr3 \citep[]{2021Gaia} distance and ZTF \citep[][]{2019pasp..131.018002} multi-color light curves, the mass of J2240 could be constrained to {\bf $0.085^{+0.036}_{-0.024} M_{\odot}$}, which is lower than any known ELM WD. 
Testing the lower limit of ELM WD is important for studying the theories of mass transferring (MT) and accreting process, 
the Common Envelope phenomena (CE),
angular momentum loss (AML) due to magnetic braking and Gravitational wave radiation, radiation driven evaporation,
and many other physical mechanisms related to close binary evolution 
\citep[][]{1986ApJ...311..742I, 2014A&A...571A..45I, 2017MNRAS.467.1874C, 2018ApJ...858...14S, 2019ApJ...871..148L}.

This paper is organized as follows.
In Section \ref{sec.obs} the existing data and new observations are gathered and described.
In Section \ref{sec.results} the data analyzing methods and results are illustrated.
The summaries and discussions are shown in Section \ref{sec.sum}.

\section{Observations}
\label{sec.obs}

In this Section, 
both archival and newly acquired observational data for J2240
are present.
For convenience through out this work,
we name the visible component as star 1, the primary,
and the invisible component as star 2, the secondary.
Phase 0 represents the moment when the visible component is at the superior conjunction point.

\subsection{Spectroscopic observation}
\label{sec:spectra}

\begin{figure*}[htbp!]
\center
\includegraphics[width=0.98\textwidth]{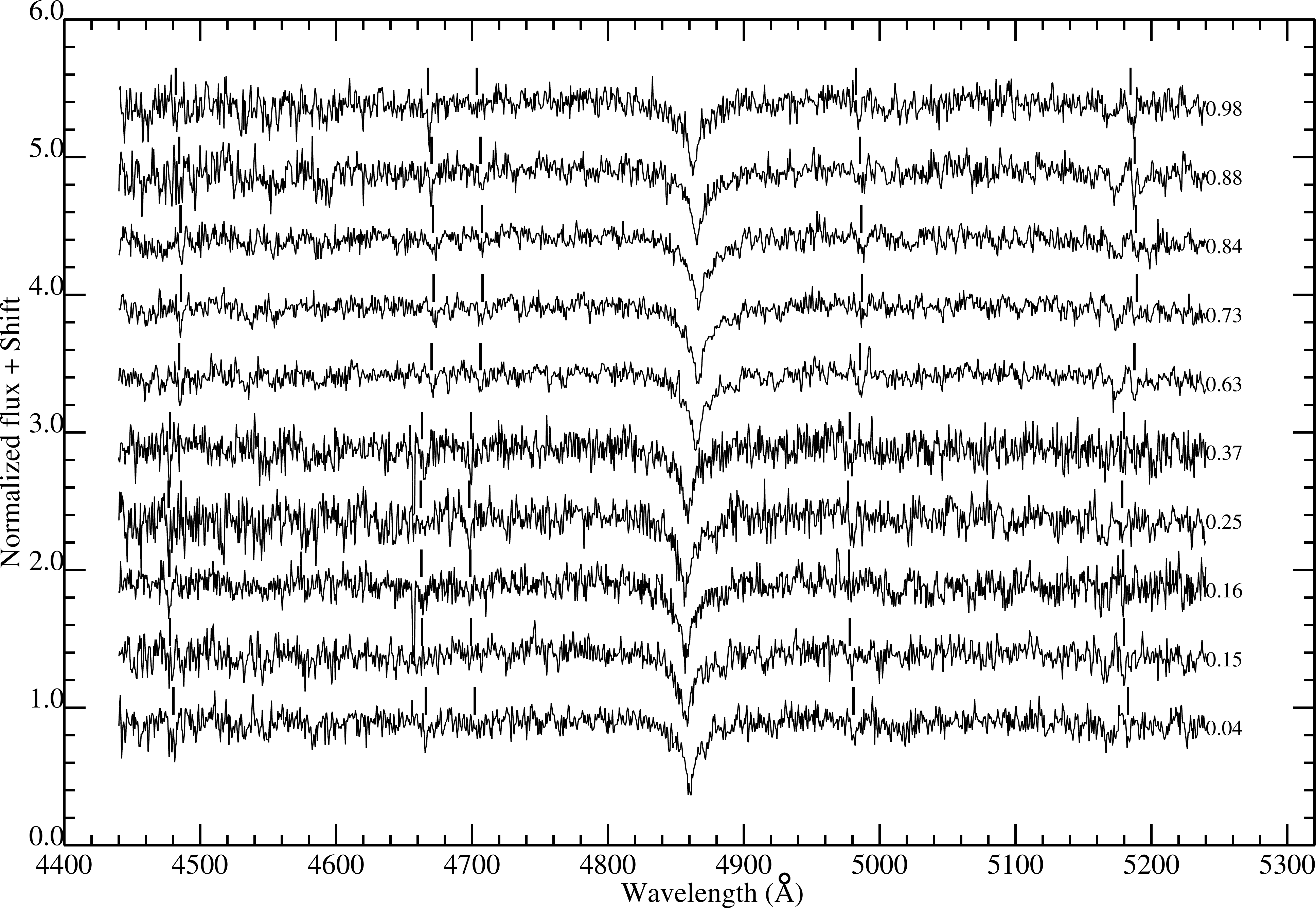}
\includegraphics[width=0.98\textwidth]{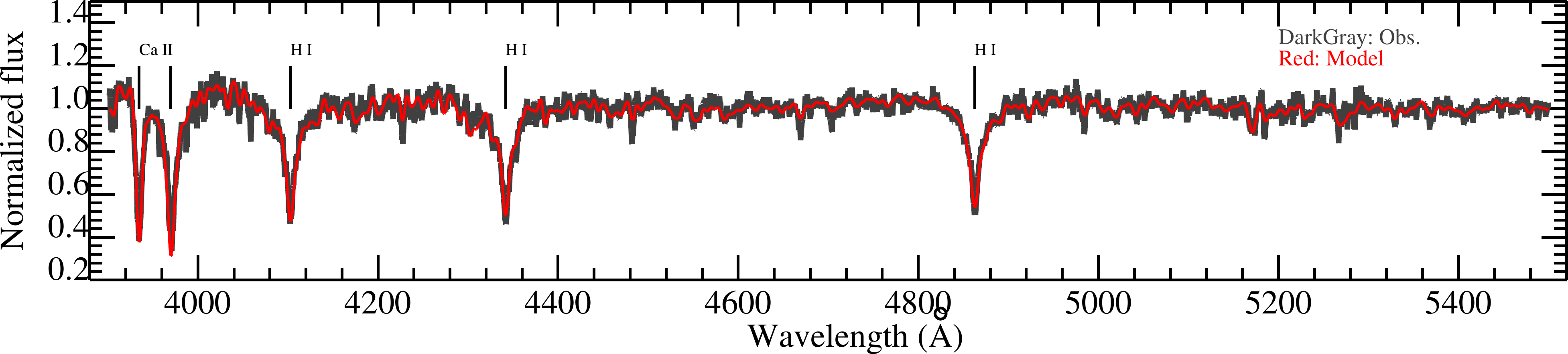}
\includegraphics[width=0.98\textwidth]{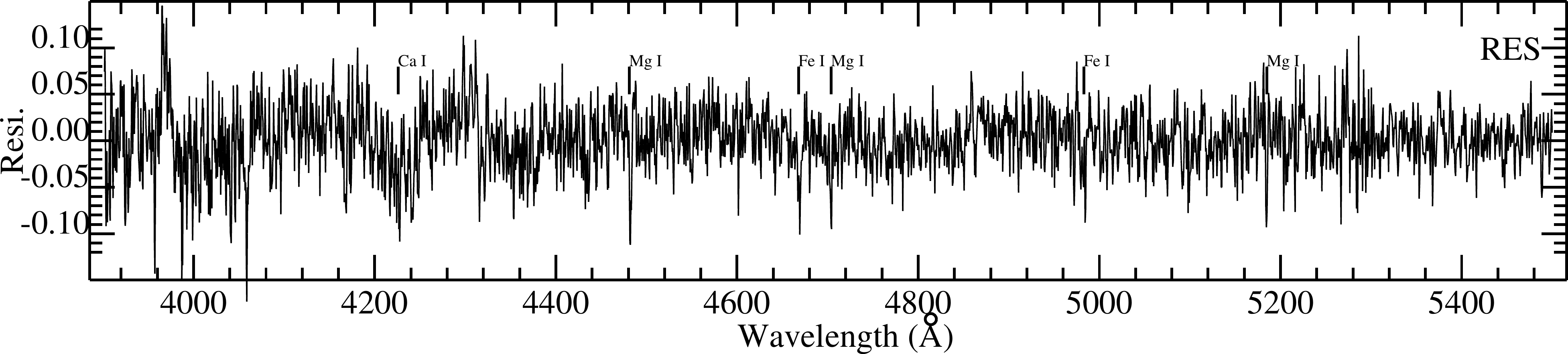}
\caption{
The upper panel: the single exposure spectra of J2240 from 
LAMOST and P200.
The spectra from bottom to top are in phase ascending order,
and adjacent spectra are shift by 0.5 in y direction to make a clear view.
The phases are shown in the right side.
Details are shown in Table \ref{tab.sp_obs}.
The middle panel:
The combined spectrum of the blue band (black),
over-plotted with its best fitting template (red).
The lower panel:
The residual in the fitting of combined spectrum of the blue band.
}
\label{spfig}
\end{figure*}

J2240 was first discovered as a candidate of binaries with
compact object in LAMOST low resolution(R $\approx$ 1800) survey \citep{2012RAA....12.1197C,2012RAA....12..723Z}.
In regular data release, the spectra of sub-exposures are combined to remove cosmic rays. To take full advantage of LAMOST spectra in time domain,
the sub-exposure spectra as well as RV measurement of individual spectrum were released in 2021 \citep[][]{2021RAA....21..249B}.
J2240 has 8 individual low resolution spectra 
in 3 nights from 2016 to 2021, 
as shown in Fig. \ref{spfig} and Table \ref{tab.sp_obs}. 
The LAMOST spectra show {\bf an} RV variation of more than 560km/s, indicating the system should be a close binary with possible compact companion. 

Two additional low resolution spectra were acquired 
in the Double Spectrograph (DBSP) of 
the 200 inch Hale telescope at Palomar Observatory(P200) via the TAP (Telescope Access Program for China based astronomers) \footnote{http://info.bao.ac.cn/tap/} project.
The dichroic filter was set to $D\-55$. 
For the blue side, 
the grating was set to 1200 lines/mm, 5000\AA\ blaze,
and the angle was set to $35^\circ52^\prime$
(center wavelength: 4500\AA).
For the red side,
the grating was set to 1200 lines/mm, 7100 $\AA$ blaze,
and the angle was set to $42^\circ02^\prime$
(center wavelength: 6600\AA).
The slit width was set to $1.5^{\prime\prime}$,
according to the seeing.
The spectra cover wavelength from 3900 to 5250\AA\ in the blue,
and from 5800 to 7400\AA\ in the red,
with a resolution $R$ $\approx2100$ and $\approx3200$, for the blue and red respectively.
The IRAF package is used to handle raw data processing, 
including bias and flat correction, cosmic ray removal, 
spectral extraction and wavelength calibration.
The template cross-correlation method used in \cite{2021RAA....21..249B} was applied to measure the RV of P200 spectra. 
The method shifts and compares theoretical templates to the observed spectra, finds the best RV at the $\chi^2$ minima, and estimates the uncertainties at $\Delta\chi^2 =1$.
Observation information as well the measured spectral parameters are shown in Table \ref{tab.sp_obs}.

\begin{table*}
\caption{Spectroscopic observations and estimated parameters of J2240.
Note the DATE column represents the time zone observation night,
and the HJD represents the heliocentric time at the middle of exposure.
The signal to noise ratios ({\bf S/N}) for single exposures are between 10 and 20, the typical errors in this {\bf S/N} range are 150k, 0.25dex and 0.15dex for $T_{\rm eff}$, log $g$ and [Fe/H], respectively.
The RVs denote the radial velocities directly measured from spectra,
while the \textbf{RVCorr}s denotes the phase smearing corrected velocities, as discussed in Section \ref{sec:orbit}.
\label{tab.sp_obs}}
\setlength{\tabcolsep}{5.0pt}
\begin{center}
\begin{tabular}{cccccccccc}
\hline\noalign{\smallskip}
DATE & HJD & RV & Phase & RVCorr & $T_{\rm eff}$ & log $g$ & [Fe/H] & EXPTIME & Tel.$/$Ins. \\
     & day & km/s  &     & km/s   & K             & dex     & dex    & s & \\
\hline\noalign{\smallskip} 
20161128 & 2457720.9407 &  141.98$\pm$4.49  & 0.62 &  145.16 & 7347 & 4.14 &  0.11 &1800 & LAMOST$/$LRS \\
20161128 & 2457720.9636 &  236.38$\pm$4.86  & 0.73 &  240.97 & 7357 & 4.08 &  0.15 &1800 & LAMOST$/$LRS \\
20161128 & 2457720.9872 &  209.65$\pm$4.42  & 0.83 &  213.84 & 7323 & 4.05 &  0.14 &1800 & LAMOST$/$LRS \\
20171120 & 2458077.9763 & -132.78$\pm$7.92  & 0.04 & -133.71 & 7334 & 4.25 & -0.11 &1800 & LAMOST$/$LRS \\
20171120 & 2458077.9992 & -303.82$\pm$6.74  & 0.14 & -307.30 & 7608 & 4.37 & -0.06 &1800 & LAMOST$/$LRS \\
20171120 & 2458078.0228 & -374.01$\pm$10.12 & 0.25 & -378.54 & 7478 & 4.36 &  0.11 &1800 & LAMOST$/$LRS \\
20210603 & 2459368.9314 & -337.18$\pm$3.34  & 0.15 & -338.94 & 7525 & 4.14 &  0.06 &1200 &  P200$/$DBSP \\
20210603 & 2459368.9787 & -299.28$\pm$4.09  & 0.37 & -300.79 & 7197 & 4.07 &  0.01 &1200 &  P200$/$DBSP \\
20211010 & 2459498.0289 & 148.55$\pm$7.30   & 0.87 &  151.83 & 7401 & 4.11 &  0.09 &1800 & LAMOST$/$LRS \\
20211010 & 2459498.0512 & -24.90$\pm$7.89   & 0.97 &  -24.21 & 7099 & 3.94 &  0.25 &1800 & LAMOST$/$LRS \\
\noalign{\smallskip}\hline
\end{tabular}
\end{center}
\smallskip
\end{table*}

\subsection{Light Curves}
\label{sec:lightcurve}

J2240 was firstly observed by the Catalina survey 
between 2007 and 2012,
and classified as ellipsoidal variable
with a period of 0.2196580 days and an amplitude of 0.24mag 
\citep{2014ApJS..213....9D}.
Between 2012 and 2018, 
ASAS-SN sparsely observed this target in V band for more than a few hundred times,
but no classification is available since it is too faint 
\citep[][]{2019MNRAS.486.1907J}.
%
The ATLAS survey (Asteroid Terrestrial-impact Last Alert System),
operated since 2015 using cyan and orange filters, 
classified this target as \textbf{NSINE}, 
indicating sinusoidal variables with large residual noise 
\citep{2018AJ....156..241H}.
The ZTF survey,
operated since 2018 using g, r and i bands, 
classified this target as \textbf{EW} (W Ursae Majoris type variables),
with a period of 0.219655 days in g band and 0.2196582 days in r band 
\citep[][]{2020ApJS..249...18C}.

\begin{figure}[htbp!]
\center
\includegraphics[width=0.48\textwidth]{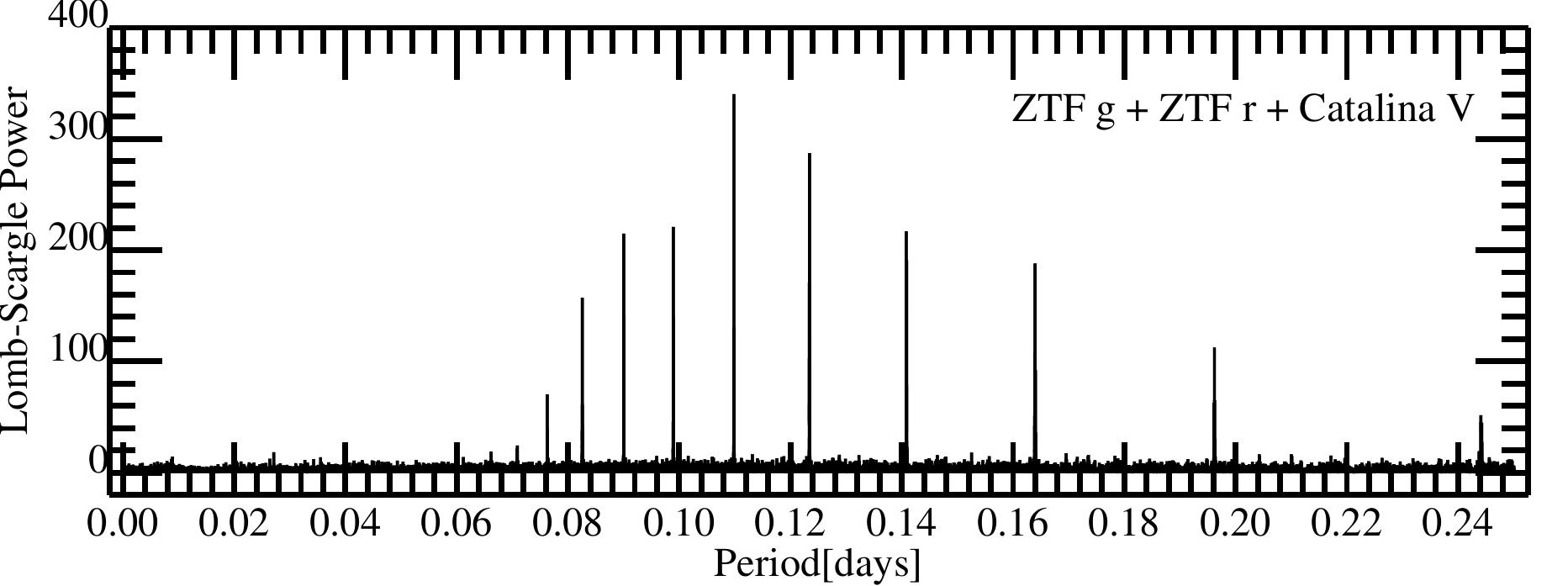}
\includegraphics[width=0.48\textwidth]{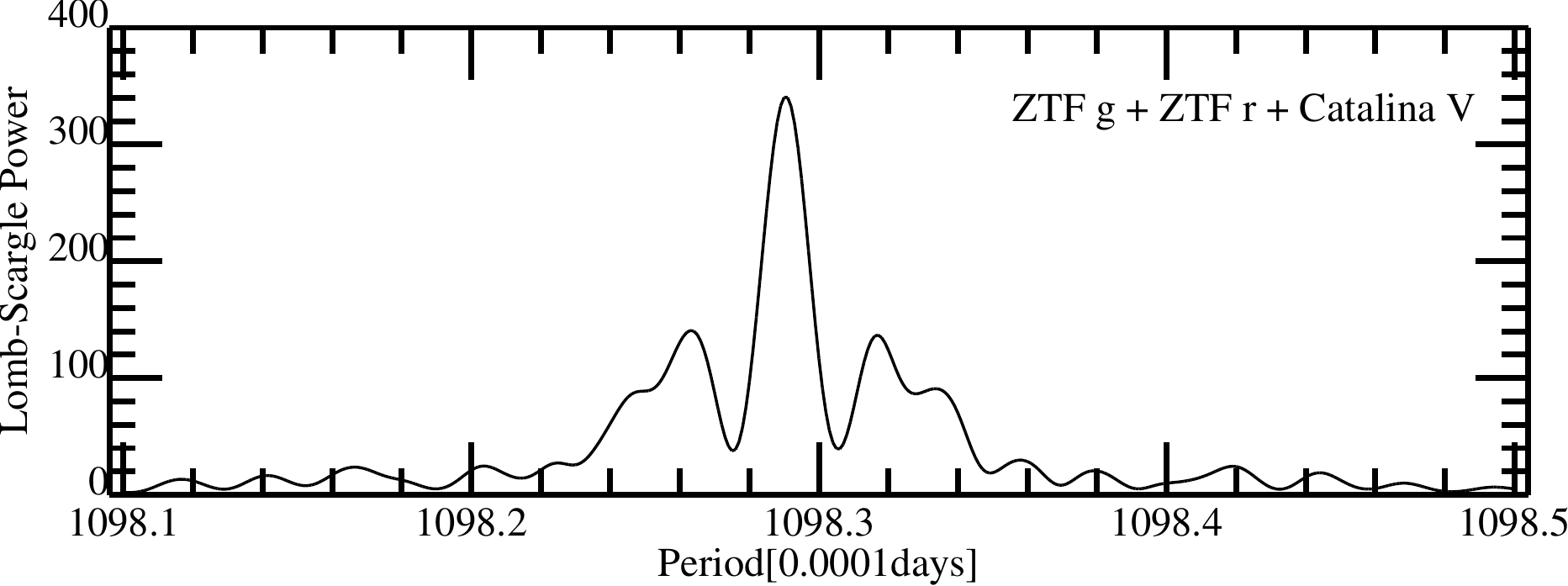}
\caption{
The Lomb-Scargle Power for the combined light curves of ZTF g, ZTF r and Catalina V.
The power peaks at half orbital period (0.109829 days), with $FWHM\approx0.000002 days.$
}
\label{fig.ls}
\end{figure}

Combing the Catalina and ZTF data (g and r band) spanning 14 years, 
we derive a period of 0.219658 days using the Lomb-Scargle periodogram \citep{1976Ap&SS..39..447L, 1982ApJ...263..835S}.
Each light curve from different bands is subtracted with the mean magnitude, divided by the standard deviation. Then curves from different bands are merged as one long span light curve.
Note the power spectrum peak is found at the half period (0.109829 days), as shown in Fig. \ref{fig.ls}.
The uncertainty, defined at the power spectrum full width half maximum ($FWHM/2.355$), is $\sim$2e-6 days.
The Automatic Fourier Decomposition \textbf{(AFD)} \citep{2015MNRAS.446.2251T} method is also used for comparison.
Given a period, the light curves are fitted to sinusoidal harmonic function (up to 6 harmonics).
The reduced $\chi^2$ values are calculated considering the residuals from multiple light curves.
The best period is found at 0.219658 via minimization of the reduced $\chi^2$, 
and the uncertainty, determined when the reduced $\chi^2$ increases by 1, is close to the value from Lomb-Scargle periodogram.

Once the period is known, by folding the light curves and fitting the points near the minimum ($\pm0.1$ phase) with sinusoidal function, the zero point can be estimated. 
The uncertainty is estimated using Markov Chain Monte Carlo (MCMC) method.
However the negative correlation between period and zero point is evident,
with a 1e-6 days offset in period corresponding to about -30 minutes offset in zero point.
Follow up continuous high time resolution monitoring could constrain the zero point with uncertainty below one minute.
Here the adopted ephemeris of the system is,
\begin{equation}
T (\phi = 0) = 2454000.0173(400)HJD + 0.219658(2) \times N,
\end{equation}
where $\phi =$ 0 corresponds to the visible component in superior conjunction, 
and $HJD$ is the heliocentric Julian Date {\bf (in UTC)}.

 \begin{figure}[htbp!]
\center
\includegraphics[width=0.48\textwidth]{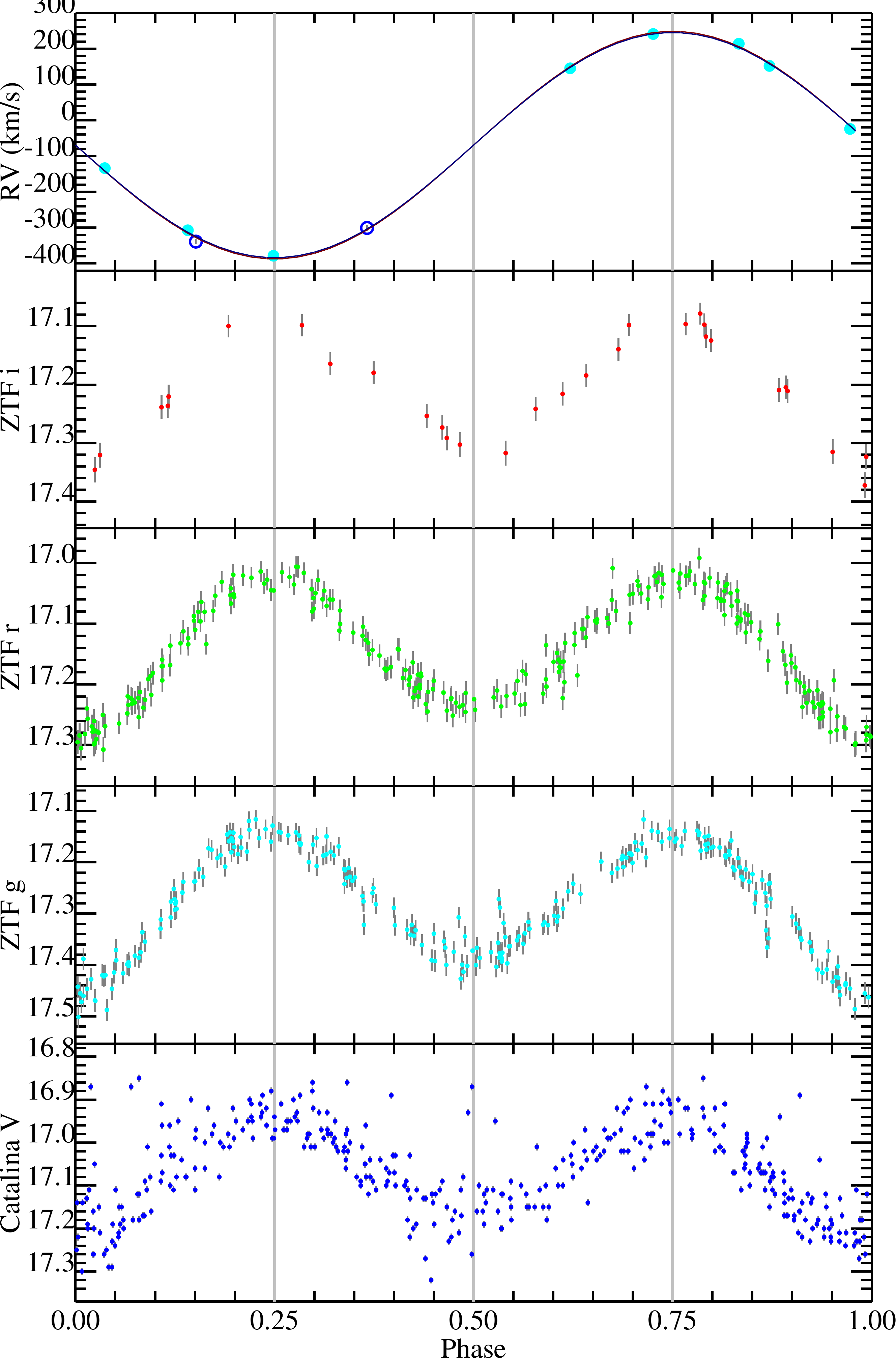}
\caption{Folded RV curves and LCs in different bands 
($V$ band of Catalina,
$g$, $r$ and $i$ bandS of ZTF)
with a period of 0.219658 day. 
The phases of 
superior conjunction ($\phi$ = 0 or 1), 
quadrature ($\phi =$ 0.25 and 0.75), 
and inferior conjunction $\phi =$ 0.5 
are marked with vertical gray lines. 
A calculated RV curve is shown 
using best fit K1$=318.5km/s$ and V$_{sys}=-68.6km/s$.
}
\label{show.fig}
\end{figure}

\subsection{Multi-band photometry}
\label{sec:photometry}

According to the GALEX Search Tool \dataset[10.17909/T9H59D]{https://doi.org/10.17909/T9H59D}
\footnote{http://galex.stsci.edu/GR6/?page=mastform},
J2240 has two records in GALEX archive \citep{2017ApJS..230...24B},
 $OBJID=$ $6376439803911605972$ and $6376439771699347883$.
For both records, only $NUV$ band measurements are valid,
which are $20.66\pm0.16$ and $20.24\pm0.15$, in magnitude, respectively.
Via CasJob \footnote{http://galex.stsci.edu/casjobs/},
a joint query between $VisitPhotoObjAll$, $VisitPhotoExtract$ and $visitImg$
shows that the target was visited 4 times (see Table \ref{tab.galex}).
The latest two visits merge into the combined record of $OBJID=6376439803911605972$,
while the two in 2004 merge into the combined record of $OBJID=6376439771699347883$.
One of the four visits has reported an FUV calibrated flux using NUV position, 
$3.99\pm4.17 microJy$ ($22.40\pm1.14$ in mag), though there is little signal in the combined image by visual inspection.
This value has apparently large error, thus could be the upper limit of FUV flux.

\begin{table}
\caption{The GALEX observation of J2240. Only NUV measurements are available.
\label{tab.galex}}
\setlength{\tabcolsep}{4.5pt}
\begin{center}
 \begin{tabular}{cccc}
\hline\noalign{\smallskip}
DATE & JD(day) & NUV mag & PHASE \\
\hline\noalign{\smallskip}
20070908 & 2454351.9646 & 20.89$\pm$0.27 & 0.24 \\
20061012 & 2454021.1541 & 20.46$\pm$0.20 & 0.22 \\
20040810 & 2453231.9007 & 20.29$\pm$0.20 & 0.11 \\
20040810 & 2453231.9692 & 20.13$\pm$0.21 & 0.42 \\
\noalign{\smallskip}\hline
\end{tabular}
\end{center}
\smallskip
\end{table}

No X-ray catalogues match J2240 directly according to \textbf{CDS Portal} 
\footnote{http://cdsportal.u-strasbg.fr//}.
The upper limit is checked via the HIgh-energy LIght-curve GeneraTor
\footnote{http://xmmuls.esac.esa.int/upperlimitserver/}.
The \textbf{XML-NEWTON SLEW} mission provides upper limits 
in the $0.2-2$, $2-12$ and $0.2-12$ flux bands
(7.1073e-13, 5.7056e-12 and 1.4153e-12 $erg/s/cm^2$ respectively, on May 16, 2009).
The \textbf{ROSAT-SURVEY} provides an upper limit of 3.5579e-13 $erg/s/cm^2$
in $0.2-2$ flux bands on Nov. 17, 1990.
The \textbf{INTEGRAL} mission provides upper limits 
in the $20-40$, $40-60$ and $60-100$ flux bands
(1.0381e-11, 8.9913e-12 and 9.4841e-12 $erg/s/cm^2$ respectively, on Nov. 7 2003).
The upper limit of X-ray to optical flux ratio 
$\log(f_X/f_V)$ can be estimated as $\approx-0.24$,
which is within the chromosphere active range 
\citep[][]{2019RAA....19...98H}.

In the near infrared band,
J2240 was visited 
by the Two Micron All Sky Survey on Sep. 29 1998 \citep[][]{2006AJ....131.1163S},
corresponding to a phase of 0.097,
with mag J $=16.75\pm0.13$,
mag H $=16.22$ and mag K $=15.94$.
However the related $Qflg$ is $BUU$ and $Rflag$ is $200$,
indicating no source is detected in H and K band
and the upper limit is adopted.
Similarly in the WISE observations,
the $qph$ flat is $ACUU$, indicating 
only the $W1$ band is reliable.
Fortunately the VISTA Hemisphere Survey (VHS) \citep[][]{2013Msngr.154...35M}
 provide better Y, J, H and Ks magnitudes.
All their bit wise processing warning/error flags are 0,
indicating reliable measurements.
So we discard the unreliable photometry, leave VISTA and WISE W1 for the following analysis.
The multi-band photometry measurements are collected in Table \ref{tab.mags}.

\begin{table*}
\caption{Multi-band photometry measurements for J2240. \label{tab.mags}}
\setlength{\tabcolsep}{4.5pt}
\begin{center}
\begin{threeparttable}
\begin{tabular}{cccc}
\hline\noalign{\smallskip}
Filter & Magnitude & Reference \\
\hline\noalign{\smallskip}
GALEX NUV & 20.6582$\pm$0.1643\tnote{1} & \citet[][]{2017ApJS..230...24B} \\
SDSS u & 18.441$\pm$0.018 & \citet[][]{2020ApJS..249....3A} \\
SDSS g & 17.350$\pm$0.005 & \citet[][]{2020ApJS..249....3A} \\
SDSS r & 17.186$\pm$0.006 & \citet[][]{2020ApJS..249....3A} \\
SDSS i & 17.195$\pm$0.006 & \citet[][]{2020ApJS..249....3A} \\
SDSS z & 17.280$\pm$0.014 & \citet[][]{2020ApJS..249....3A} \\
PS1 g & 17.3403$\pm$0.0529 & \citet[][]{2016arXiv161205560C} \\
PS1 r & 17.2061$\pm$0.0059 & \citet[][]{2016arXiv161205560C} \\
PS1 i & 17.2487$\pm$0.0133 & \citet[][]{2016arXiv161205560C} \\
PS1 z & 17.2694$\pm$0.0162 & \citet[][]{2016arXiv161205560C} \\
PS1 y & 17.3129$\pm$0.0222 & \citet[][]{2016arXiv161205560C} \\
GaiaEDR3 G & 17.1381$\pm$0.0073 & \citet[]{2021Gaia} \\
GaiaEDR3 BP & 17.2973$\pm$0.0240 & \citet[]{2021Gaia} \\
GaiaEDR3 RP & 16.8414$\pm$0.0206 & \citet[]{2021Gaia} \\
2MASS J	& 16.754$\pm$0.134 & \citet[][]{2003cvs..book.....W} \\
2MASS H & 16.221\tnote{2} & \citet[][]{2003cvs..book.....W} \\
2MASS Ks & 15.941\tnote{2} & \citet[][]{2003cvs..book.....W} \\
WISE W1 & 16.428$\pm$0.079 & \citet[][]{2014yCat.2328....0C} \\
WISE W2 & 16.998$\pm$0.457\tnote{3} & \citet[][]{2014yCat.2328....0C} \\
WISE W3 & 12.302\tnote{3} & \citet[][]{2014yCat.2328....0C} \\
WISE W4 & 8.722\tnote{3} & \citet[][]{2014yCat.2328....0C} \\ 
VISTA Y & 16.7193$\pm$0.0131 & \citet[][]{2013Msngr.154...35M} \\
VISTA J & 16.6129$\pm$0.0146 & \citet[][]{2013Msngr.154...35M} \\
VISTA H & 16.5509$\pm$0.0164 & \citet[][]{2013Msngr.154...35M} \\
VISTA Ks & 16.4528$\pm$0.0393 & \citet[][]{2013Msngr.154...35M} \\
\noalign{\smallskip}\hline
\end{tabular}
\begin{tablenotes}
\footnotesize
\item[1] This is the combined magnitude for $OBJID=$ $6376439803911605972$.
\item[2] This is an upper limit. 
\item[3] The photometry quality flag is not good and no error is provided for W3 and W4.
\end{tablenotes}
\end{threeparttable}
\end{center}
\end{table*}

\subsection{Distance, Kinematics and Extinction}
\label{sec:distance}

In Gaia EDR3 \citep[]{2021Gaia},
J2240 is $source\_id=2652639593773429120$.
Its parallax is $0.69426mas$ and parallax error is $0.085375mas$.
The $RUWE$ is $1.0642716$,
implying no significant excess noise 
and no indicative of parallax estimation problem.
%
%
%
The parallax bias, according to \citet[][]{2021A&A...649A...4L},
is $-0.03054mas$, yielding to a corrected parallax of $0.7248$,
and corresponding to distance of $1380^{+184}_{-145}pc$, 
and distance modulus of $10.700^{+0.272}_{-0.242}$.
The distance using Bayes technology, 
according to \citet[][]{2021AJ....161..147B},
is $1340.4^{+185.7}_{-132.3}pc$.
%
In case the effective temperature is unchanged,
a smaller distance indicates smaller stellar radius and mass estimation.
Here we choose to use the distance of $1380^{+184}_{-145}pc$
in our following work.

%
%
J2240 is within the southern Galactic cap,
and is $\sim1000pc$ below the Galactic plane according to its distance.
Its proper motion in Gaia EDR3 is $pmra=-2.6559\pm0.1004mas$ and $pmdec=0.6671\pm0.0806mas$.
Together with system velocity of $-68.6\pm2.6km/s$ from our orbit fitting (Section \ref{sec:orbit}),
the 3D motion in LSR $(U,V,W)_{LSR}$ is $(9.5\pm0.7km/s, -26.9\pm1.5kms, 58.9\pm2.0kms)$,
indicating a thin disk object 
($P_{thin}=70\pm5\%$, $P_{thick}=28\pm4\%$, $P_{halo}=2\pm1\%$)
\citep[][]{2013ApJ...764...78R}.
Here the bootstrapping method is used to estimated the uncertainties.

The 3D dust map of \citet[][]{2019ApJ...887...93G},
which is based on Gaia parallaxes and stellar photometry 
from Pan-STARRS1 and 2MASS,
gives an extinction coefficient $\alpha$ of 0.07.
Actually the coefficient $\alpha$ reaches 0.07 at distance modulus $DM=6.25$ and remains unchanged for larger $DM$s.
The corresponding extinction coefficient $E(B-V)$ is $0.062$ or $0.069$,
according to the two transformation formula of \citet[][]{2019ApJ...887...93G}.
2D extinction based on IRAS and COBE/DIRBE data shows lower value.
$E(B-V)$ from \citet[][]{1998ApJ...500..525S} and \citet[][]{2011ApJ..737103} is 0.0506$\pm$0.0006 and 0.0435 $\pm$ 0.0005 respectively. 
We here adopt $E(B-V)=0.056\pm0.013$ by choosing 0.0435 and 0.069 as the 16\% and 84\% percentiles,
Then we use it as a prior in the Spectral Energy Distribution (SED) fitting in Section \ref{sec:sedfit}.
The returned posterior extinction, 0.055$\pm$0.013 (see Figure \ref{fig.sedfit_mcmc}), is well consistent with the prior.
Based on the extinction law of \citet[][]{1989ApJ...345..245C} 
and the band effective wavelength adopted from the Wilson-Devinney code
\citep{1971ApJ...166..605W},
the corresponding extinctions in g, r, i and Gaia G bands are 
0.204$\pm$0.048, 0.149$\pm$0.035, 0.112$\pm$0.026 and 0.152$\pm$0.036, respectively.

%
%
%

\section{Methods and Results}
\label{sec.results}

In Section \ref{sec:spparam}, we estimate spectral atmosphere parameters using template cross match method.
In Section \ref{sec:sedfit}, SED fitting is carried out using MCMC technique.
With observed magnitudes, extinction, distance and effectively temperature, 
the radius can be estimated properly, with distance contributing majority of the uncertainty.
In the following orbital parameter fit with RV curve in Section \ref{sec:orbit}, 
we obtain RV semi-major amplitude of the visible component, 
and hence the binary mass function of the invisible component.

In Section \ref{sec:wdfit} the light curves from ZTF are fitted simultaneously with RV data.
The ellipsoidal amplitude ($\approx30\%$) provides constrain on the filling factor of the visible component, the orbital inclination and also the mass ratio \citep{2021MNRAS.501.2822G}. 
The visible component of our binary has small radius but relative large filling factor, suggesting the orbital separation is small.
As the orbital period is strongly constrained, the small separation thus indicates small masses for both components.
In the absence of light curve eclipse feature, the inclination angle is weakly constrained,
as it has a strong negative correlation with the filling factor, which directly determines the amplitude of the tidal deformation.
Nevertheless, the masses of both component can still be properly constrained into a small range.

The Wilson-Devinney code \citep[][]{1971ApJ...166..605W} is a well performed and widely used code for simulating binary light curves and radial velocity curves.
The code provides two executable program, the \textbf{Light Curve} (LC) and Differential Correction (DC).
The first one (LC program) computes light curves from different filters, given necessary physical parameters of the binary systems.
The second one (DC program) calculates the numerical derivative and finds the local optimum using Levenberg-Marquardt algorithm.
Among several tens of variables, which can be directly given and adjusted in the Wilson-Devinney code, 
effective temperatures, {\bf Roche} potential (related to radius), mass ratio, inclination angle and separation are of the most critical importance.
While mass, radius, luminosity and surface gravity of each component can be only be calculated indirectly, by LC program or by external code.
We \textbf{preliminarily} try the Wilson-Devinney code to constrain the system (inclination, orbital separation, mass ratio and {\bf Roche} potential) without change distance and extinction.
Then we derive the uncertainties by considering the uncertainties of distance, extinction and other related parameters, with \textbf{bootstrapping} method.
Details are described step by step in Section \ref{sec:wdfit}.

\subsection{Spectral parameters}
\label{sec:spparam}
As the magnitude of our binary is relatively faint,
the {\bf S/N} of both the LAMOST and P200 spectra are between 10 and 20(see Fig\ref{spfig}).
The spectra
are dominated by
plateau Ca II H $\&$ K in absorption, 
normal Balmer series, 
weak Mg I triplets and Ca II Triplets,
nearly invisible G band, Ca I 4226 line and Paschen series,
suggesting our target is an early F-type star.
All of these lines are in absorption.
Helium lines are invisible in all the spectra.
The UlySS package \citep[][]{2009A&A...501.1269K}, which is adopted in LAMOST stellar parameterization pipeline \citep{2014IAUS..306..340W},
is used to match the low resolution spectra to the ELODIE high resolution models to derive the stellar parameters.
In the range of 3830 and 5500 \AA, the high resolution model spectra are radial velocity shifted, degraded into low resolution by convolution with a Gaussian kernel,
and multiplied with an auto adjusted polynomial to act as flux scaling/normalization.
The best fitting parameters are then found by minimizing the $\chi^2$ value, while the uncertainties are determined when $\Delta\chi^2=1$.
The parameters for each single exposure are listed in Table \ref{tab.sp_obs}.
Since the {\bf S/N} of each individual spectrum is relatively low, 
 all the single exposures were shifted to the rest frame and combined to get a spectrum with higher {\bf S/N}. The stellar parameters derived from the rest frame co-added spectrum are $T_{\rm eff}=7198\pm150K$,
log $g$=4.29$\pm$0.25dex and [Fe/H]=0.08$\pm$0.15dex
\footnote{As the error given by UlySS package is unrealistically small, the typical measurement error from https://www.lamost.org/dr8/v1.1/doc/release-note are adopted.},
which agrees with with the average of the individual measurement,i.e., $T_{\rm eff}=7361\pm139K$,
log $g$=4.14$\pm$0.13dex and [Fe/H]=0.08$\pm$0.12dex.
The adopted spectral parameters are shown in Table \ref{tab.wdfit}.

The best fit model and the residuals are shown
in the middle and bottom panel of Fig. \ref{spfig}, respectively.
As marked in the figure, there are a few lines that are 
slightly deeper than the model in the residual spectrum.
As shown in the upper panel of Fig. \ref{spfig}, 
these absorption features move in phase with the visible component, suggesting they are not interstellar absorption lines.
Considering the low resolution and relative low {\bf S/N} of individual spectra,
further high resolution spectroscopic observation will be useful 
to verify the stellar parameters and element abundance,
and then better understand the evolution stage of the system. 

\subsection{Spectral Energy Distribution fitting}
\label{sec:sedfit}

Since the photometry varies with the phase by up to 30\%,
the phase averaged magnitudes should be used with higher priority to obtain a good SED fit. 
Considering our target is in a close binary system, interaction may change their evolution path, 
thus we only fit the SED with KURUCZ \citep{Castelli2003} spectral template to derive the temperature and radius, rather than fitting with stellar evolution model.
The model flux is calculated by convolving the filter transmission function with the model spectra.
While the observations are obtained via VizieR Photometry Viewer \footnote{https://vizier.cfa.harvard.edu/vizier/sed/}, where data are already transformed into the unit of \textbf{Jansky}.

J2240 was observed by Pan-STARRS (PS1) for 
14, 22, 20, 15 and 12 times in g, r, i, z and y bands, 
respectively.
Certainly, the sampling time are not evenly distributed within the period,
and the epochs for different bands are different.
With model light curves from preliminary fitting and 
the real observation phases from PS1 survey,
we can calibrate the PS1 magnitude to the average magnitude of the light curve of the corresponding filter.
However, due to the relative large uncertainty of the light curve period and zero point,
the uncertainties of corrections are comparable to the corrections.
Besides preliminary tests show that the influence of these corrections is tiny ($\sim30K$ in temperature), {\bf therefore} we decide not to over correct the data.
In the infrared band, we use VISTA data and WISE W1 band.
In the UV band, GALEX NUV is our only choice.
The SDSS u band is added to increase the number of points in the blue band.
{\bf To account for the phase variation, we simply enlarge the errors of SDSS u, VISTA J, H, Ks, and WISE W1 data by 3 times,
so the errors ($\sim5\%$) are comparable to the light curve amplitude}.

\begin{figure}[htbp!]
\center
\includegraphics[width=0.45\textwidth]{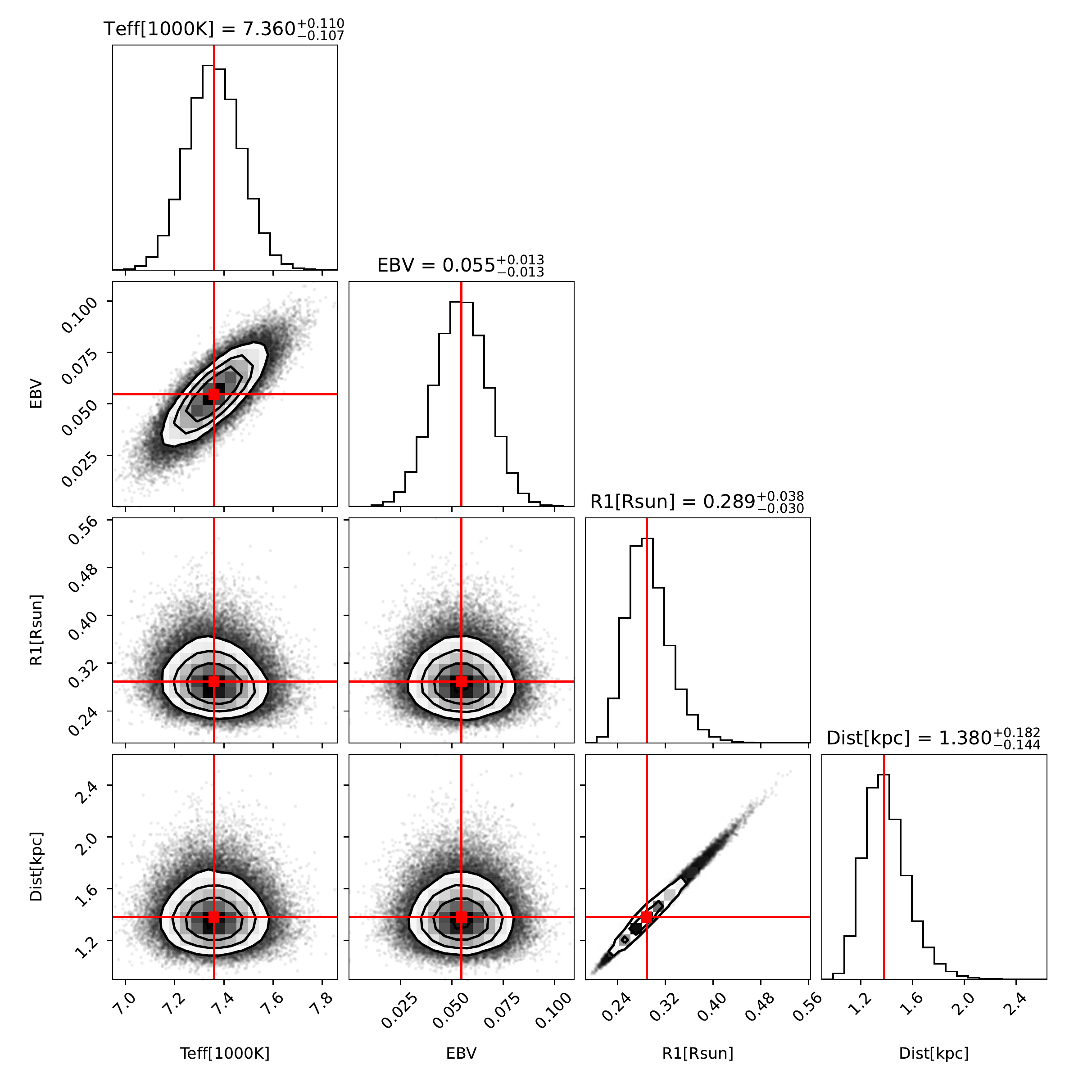}
\caption{
The corner plot of SED fitting results of J2240 with single component.
The KURUCZ model is used, 
while observations include GALEX, SDSS u, PS1, VISTA, and WISE W1.
See text for details.
}
\label{fig.sedfit_mcmc}
\end{figure}
\begin{figure}[htbp!]
\center
\includegraphics[width=0.47\textwidth]{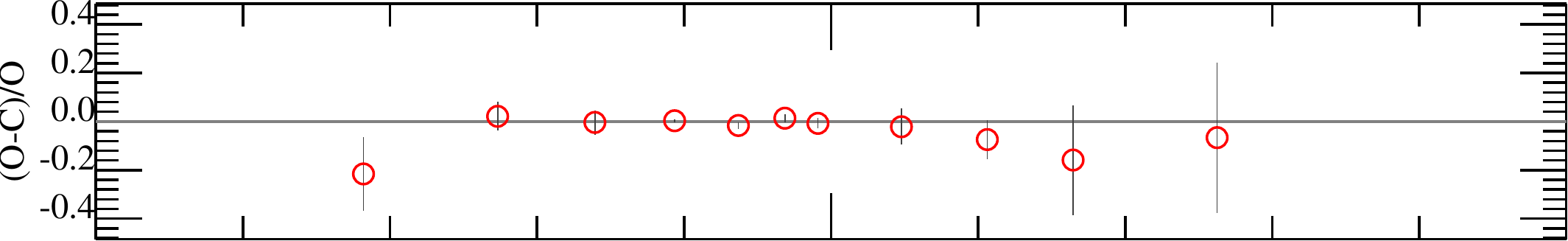}
\includegraphics[width=0.47\textwidth]{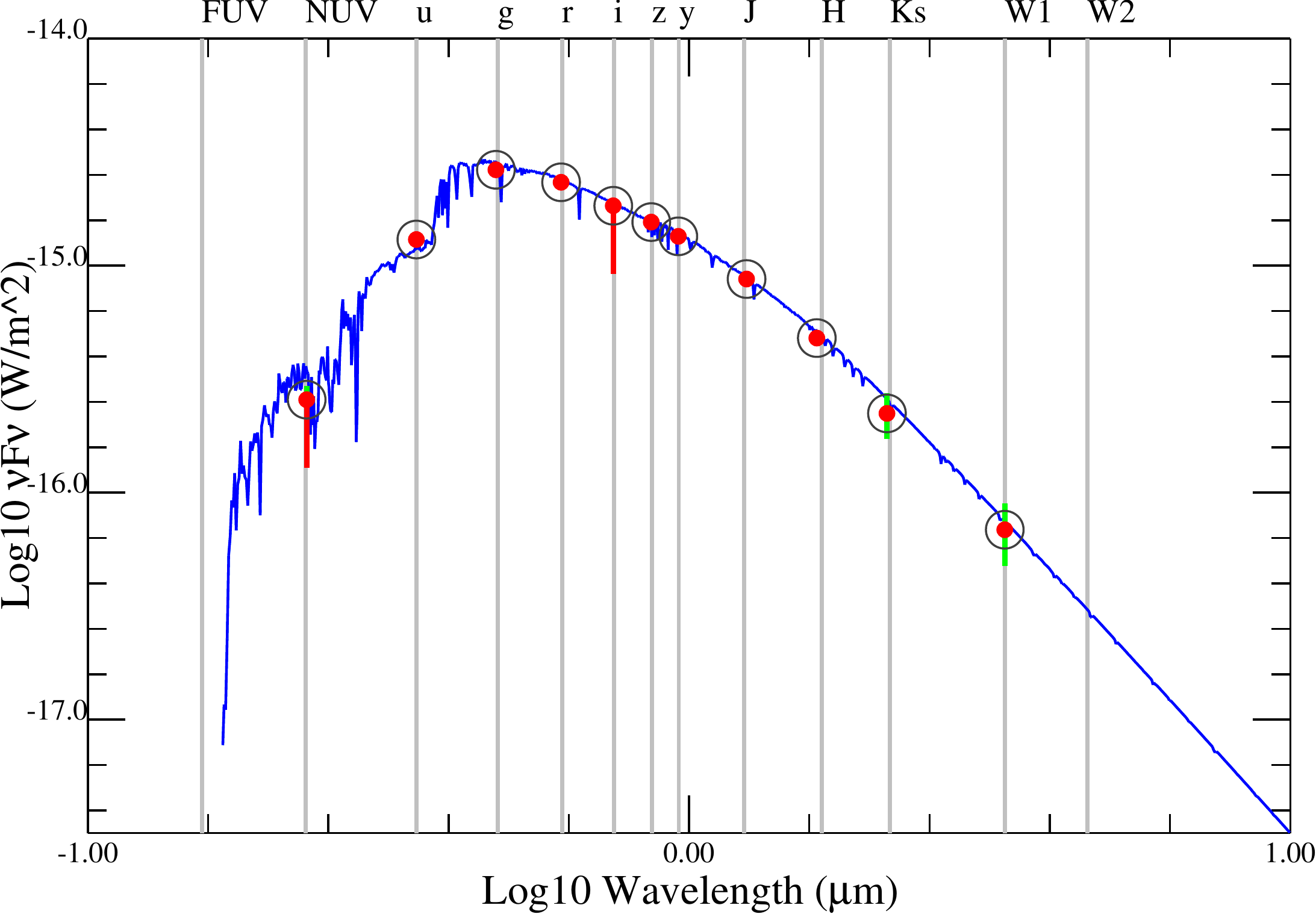}
\caption{
The the SED fitting of J2240 with single component. 
The red dots denote observations used for fitting,
including GALEX, SDSS u, the PS1, VISTA, and WISE observations.
The model parameters are $T_{\rm eff}=7360K$, log $g=4.43$ and [Fe/H]=0.08,
and the E(B-V) is 0.055. 
The residuals are expressed as ($(Observed - Calculated) / Observed$, 
and the vertical lines denote the error.
}
\label{fig.sedfit}
\end{figure}

Since the SED fitting is mostly sensitive to the temperature, 
we fixed log $g$=4.43 (from light curve fitting, Section \ref{sec:wdfit})
and [Fe/H]$=0.08$ (from spectral estimation).
The log $g$ from light curve fitting is adopted since 
the gravitational potential is better described by the shape of the light curve, thus the log $g$ could be better solved than from the low resolution spectrum.
We have tried to alter log $g$ to 4.15 and [Fe/H] to 0.0 respectively, the corresponding change of $T_{\rm eff}$ are $\approx20K$ and $\approx10K$, respectively.
As a result, the effects of both log $g$ and [Fe/H] on $T_{\rm eff}$ are negligible.
The MCMC technique is used here to optimize the fitting.
The extinction prior is set to 0.056$\pm0.013$ according to Section \ref{sec:distance}.
The distance from Section \ref{sec:distance} is also adopted as a prior.
The fitting result,
using PS1 magnitudes, SDSS u, GALEX NUV mag, VISTA J, H, Ks mag, and WISE W1,
is shown in Fig. \ref{fig.sedfit_mcmc}. 
The SED with the returned parameters ($T_{\rm eff}=7360\pm110K$, $E(B-V)=0.055\pm0.013$ and $R1=0.29^{+0.04}_{-0.03}R_{\odot}$) is shown in the Fig. \ref{fig.sedfit}.
We have also tested the influence of the inclusion of the GALEX NUV, SDSS u and WISE W1 data, and also the correction of PS1 data.
The temperate will change with deviation of about 40K, so we would like to enlarge the uncertainty of $T_{\rm eff}$ to 120K.

\begin{figure}[htbp!]
\center
\includegraphics[width=0.47\textwidth]{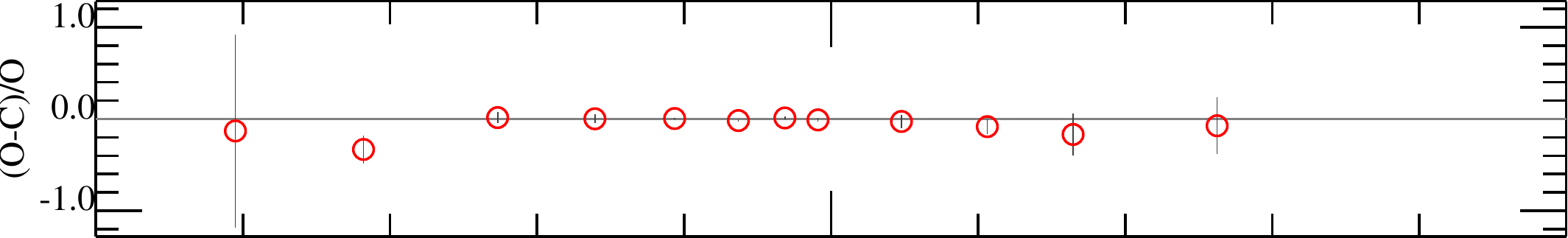}
\includegraphics[width=0.47\textwidth]{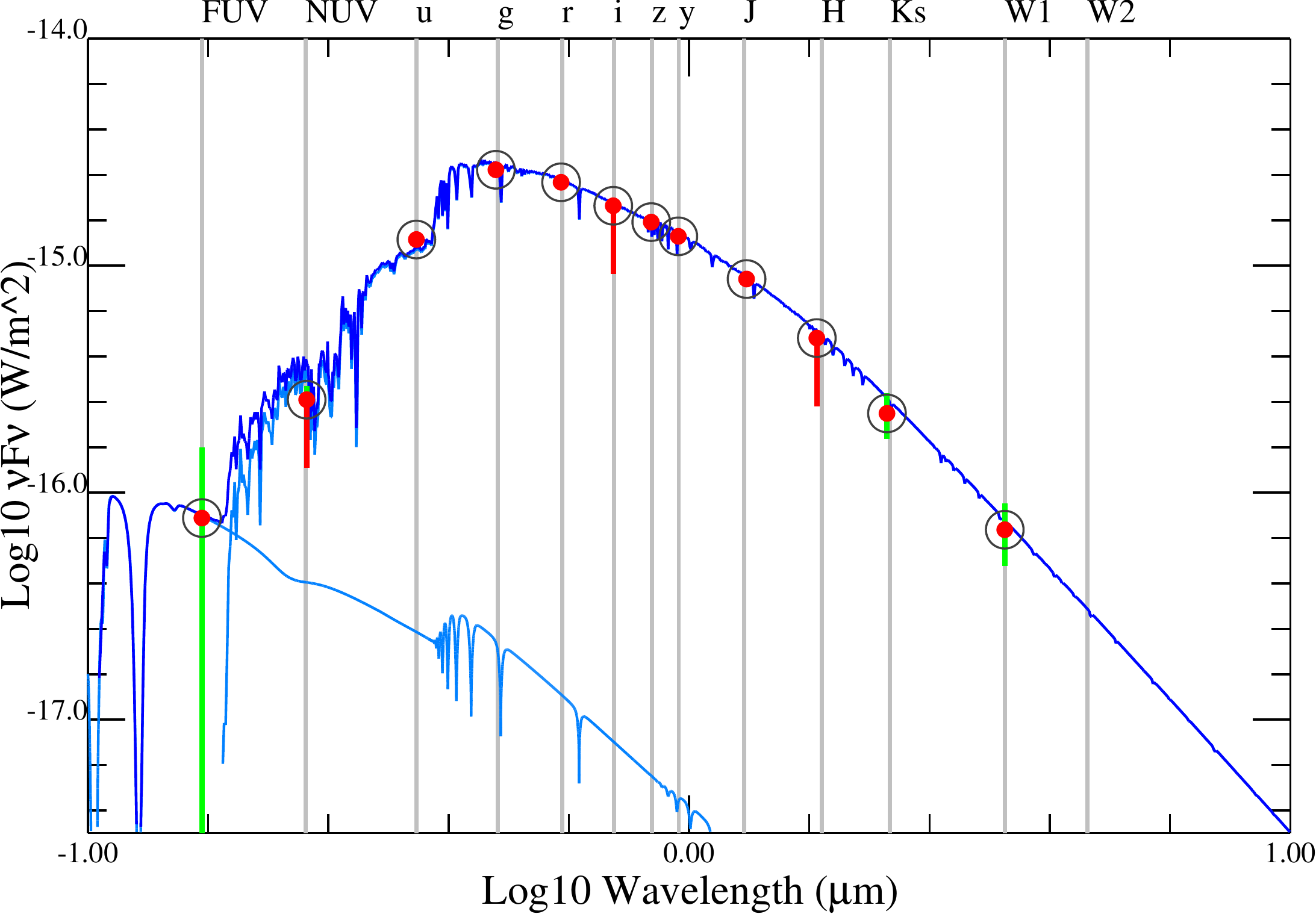}
\caption{
The SED fitting of J2240 alike Fig. \ref{fig.sedfit} with additional white dwarf component. 
The Koester WD model \citep{2010MmSAI..81..921K} is used.
The green line denotes the error of the FUV {\bf magnitude}.
The model parameters of the visible component are $T_{\rm eff}=7320K$, log $g=4.43$ and [Fe/H]=0.08,
and the E(B-V) is 0.055. 
The WD component is chosen to have $T_{\rm eff}=18000K$, log $g=7.5$ and radius $\approx0.009R_{\odot}$.
The flux contributions of the white dwarf in the optical bands are $\sim$ 1\%.
}
\label{fig.sedfitdouble}
\end{figure}

As mentioned in Section \ref{sec:photometry},
of the four single GALEX visits as shown in Table \ref{tab.galex}, one has reported a FUV calibrated flux using NUV position.
To test the flux contribution of the unseen secondary, assuming a white dwarf using Koester Model \citep{2010MmSAI..81..921K}, 
we include the FUV flux in the SED fitting.
With the constraint of FUV flux, the companion WD could be limited to cooler than $\sim$22000K, and the contributions from the hot component in the optical bands
are on the order of 1\%.
Tests also show that the inclusion of WD component has very small influence on the temperature of the visible component (about 40K).
An example is shown in Fig. \ref{fig.sedfitdouble}, with a WD temperature of 18000K.

\subsection{Orbital parameters}
\label{sec:orbit}

%
%

The RV data from both LAMOST and P200,
as shown in Table \ref{tab.sp_obs},
is used to solve the orbital parameters.
We firstly tried to fit the data use TheJoker \citep{2017ApJ...837...20P}, 
which is a well preformed Monte Carlo sampler for sparser RV data.
The returned orbital parameters are 
$p=0.2196577\pm0.0000003day$,
$e=0.02\pm0.02$, 
$K1=320.01\pm3.39km/s$ and 
$V_0=-69.14\pm2.68$.
Further more, with the eccentricity fixed to 0,
a sinusoidal curve fit is carried out via \textbf{MCMC} technique.

\begin{figure}[htbp!]
\center
\includegraphics[width=0.48\textwidth]{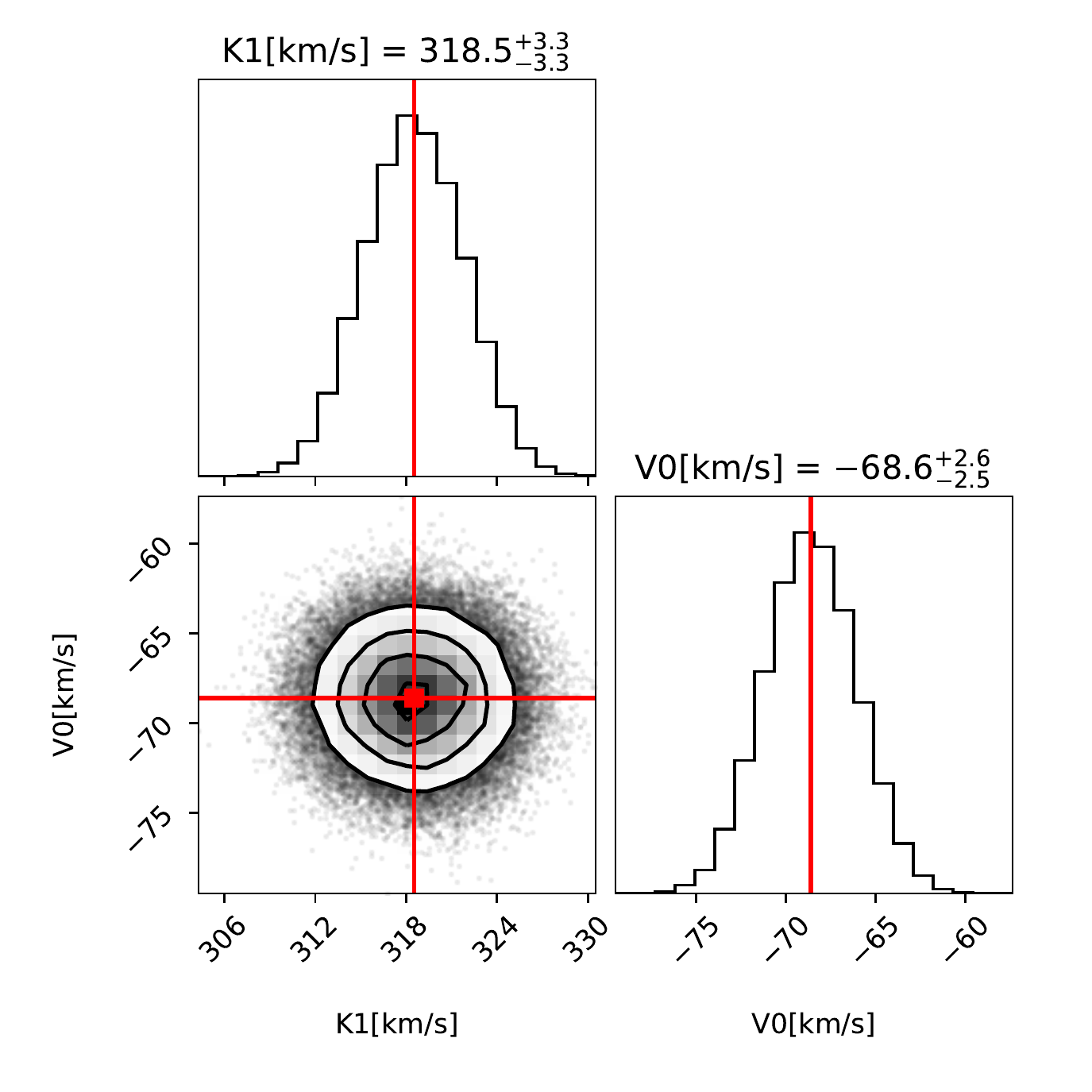}
\caption{Results of sinusoidal RV fitting using \textbf{MCMC} technique with
the eccentricity fixed to 0. 
The returned K1 is $318.5\pm3.3km/s$ 
and V0 is $-68.6\pm2.6km/s$.
}
\label{fig.rvfite0}
\end{figure}

The exposure time for spectral observation is 20-30 minutes,
which is relatively long comparing to the orbital period 5.3 hours.
The long exposure time will lead to the phase smearing effect,
which will lower the RV amplitude. 
Assuming a circular orbit, the smearing effect can be approximated by
\begin{equation}
  V(\phi)-V_0=(V_{obs}-V_0) \cdot 2\pi\delta\phi/ \sin(2\pi\delta \phi), 
 \label{eqsmear}
\end{equation}
where $V(\phi)$ is the actual velocity at phase $\phi$,
$V_{obs}$ is the measured velocity from the spectrum,
$V_0$ is system velocity of the binary and $\delta \phi$ is the
half length of the exposure time in phase space. 
For LAMOST, the half exposure length is $\delta \phi\approx1800s/2P=0.0474$ in phase,
then we have amplitude scaling factor $2\pi\delta\phi/\sin(2\pi\delta\phi)=1.015$.
As for the case of P200, the 1200s exposure time corresponds to scaling factor of 1.007.
A preliminary fitting results, using observed RV without correction, are
$K1\approx314km/s$ and $V0\approx-70km/s$.
The phase smearing corrected RV is shown 
in the `RVCorr' column of Table \ref{tab.sp_obs}.
Then we redo the fit with the corrected RV, the revised velocity amplitude
are $K1=318.5\pm3.3km/s$ and system velocity $V0=-68.6\pm2.6km/s$,
as shown in Fig. \ref{show.fig} and Fig. \ref{fig.rvfite0}.



{\bf With the estimated orbital period P and radial velocity amplitude K1, }
the binary mass function is,
\begin{equation}
  f(M) = \frac{M_{1} \, q^3 \, \textrm{sin}^3 i} {(1+q)^{2}} = \frac{P \, K_{1}^{3} \, (1-e^2)^{3/2}}{2\pi G} = 0.74M_{\odot},
  \label{eqmass}
\end{equation}
where $M_{1}$ is the mass of the primary, 
$q = M_{2}/M_{1}$ is the mass ratio, 
and $i$ is the system inclination.
Assuming an edge-on orbit and $M_{1}$=0.08$M_{\odot}$, the least mass of $M_{2}$ would be 0.88 $M_{\odot}$, which will be about 1 mag brighter than star 1 and will dominate the spectra, if it's a normal star. Thus the companion should be a compact star. i.e., a white dwarf or a neutron star. 


\subsection{Light curve fitting with Wilson-Devinney code}
\label{sec:wdfit}

\begin{table*}
\caption{Summaries of parameters for J2240.
\label{tab.wdfit}}
\setlength{\tabcolsep}{4.5pt}
\begin{center}
\begin{tabular}{cccc}
\hline\noalign{\smallskip}
Name & Description & Unit & Value \\
\hline\noalign{\smallskip}
RA & Right Ascension J2000 & [degrees] & 340.169878411 \\
DEC & Declination J2000 & [degrees] & -2.125739518 \\
P & Orbital Period & [days] & 0.219658(2) \\
T0 & Heliocentric Julian Day & [days] & 2454000.0173(400) \\
V0 & System velocity & [km/s] & $-68.6\pm2.6$ \\
K1 & RV semi-amplitude & [km/s] & $318.5\pm3.3$ \\
\hline\noalign{\smallskip}
Plx & Parallax & [mas] & $0.7248\pm0.08538$ \\
D & Distance & [pc] & $1380^{+184}_{-145}$ \\
DpcLog & $log_{10}(Distance)$ & [pc] & 3.1398 \\
E(B-V) & Reddening in B-V color & [mag] & 0.055$\pm$ 0.013 \\
Ag & Extinction in g band & [mag] & 0.204$\pm$0.048 \\
Ar & Extinction in r band & [mag] & 0.149$\pm$0.035 \\
Ai & Extinction in i band & [mag] & 0.112$\pm$0.026 \\
AG & Extinction in Gaia G band & [mag] & 0.152$\pm$0.036 \\
ALB & Bolometric albedos & [-] & $0.5-0.7$ \\
GR & Gravity Darkening exponent & [-] & $0.3-0.5$ \\
CALIB g & g band zero point & [$erg s^{-1} cm^{-3}$] & 0.49875 \\
CALIB r & r band zero point & [$erg s^{-1} cm^{-3}$] & 0.288637 \\
CALIB i & i band zero point & [$erg s^{-1} cm^{-3}$] & 0.195711 \\
\hline\noalign{\smallskip}
$T_{\rm eff}$ & Spectral temperature & [K] & 7360$\pm$150 \\
log $g$ & Spectral surface gravity & [dex] & 4.14$\pm$0.25 \\
$\left[{\rm Fe/H}\right]$ & Spectral metallicity & [dex] & 0.08$\pm$0.15 \\
$T_{\rm eff}$ & SED fitted temperature & [K] & 7360$\pm$120 \\
R1 & SED fitted star 1 radius & [$R_{\odot}$] & $0.29^{+0.04}_{-0.03}$ \\
\hline\noalign{\smallskip}
M1 & Primary mass & [$M_{\odot}$] & $0.085^{+0.036}_{-0.024}$ \\
R1 & Primary radius & [$R_{\odot}$] & $0.295^{+0.038}_{-0.030}$ \\
T1 & Light curve temperature & [K] & 7440$\pm$80 \\
log $g$ & Light curve surface gravity & dex & 4.43$\pm$0.05 \\
SMA & Semi major axis & [$R_{\odot}$] & $1.57^{+0.07}_{-0.06}$ \\
INC & Orbital Inclination & [degrees] & $60-90$ \\
M2 & Secondary mass & [$M_{\odot}$] & $0.98^{+0.16}_{-0.09}$ \\
\hline\noalign{\smallskip}
\end{tabular}
\end{center}
\smallskip
\end{table*}

As in Fig. \ref{show.fig},
the ellipsoidal variation dominants all the folded light curves.
The tidally distorted light curves offer a better constraint on the mass ratio and radius of the primary than the low resolution spectra \citep[][]{2012ApJ..74942}.
The Wilson-Devinney code \citep[][]{1971ApJ...166..605W} is used 
to model the ZTF g, r and i band light curves simultaneously.
The Wilson-Devinney code atmosphere model uses SDSS filters,
which are different from ZTF g r and i filters.
But since the ZTF magnitudes are calibrated to Pan-STARRS system \citep[][]{2019PASP..131a8003M}, 
ZTF magnitudes are converted to SDSS using the PS1 to SDSS relation \citep[][]{2012ApJ...750...99T}.
The multi-color light curves also offer an independent solution of the temperature of the visible component other than the spectra and SED fitting. 

A preliminary fit is carried out firstly.
Here, the parameters to be adjusted with \textbf{DC} program of the Wilson-Devinney code only include
the temperature of the primary $T1$,
the mass ratio $q=(M_2/M_1)$,
semi-major axis $SMA$,
inclination angle $INC$,
and the potential of the visible component $\Omega_{\rm 1}$.
For the moment distance and extinction is treated as fixed parameters and will be considered in the following bootstrapping step.
The invisible compact secondary (star 2) is modeled 
by setting a very large potential ($\Omega_{rm 2}$),
which will ensure an extremely small radius,
and temperature of 18000K from Section \ref{sec:sedfit}. 
The limb darken coefficients are internally determined 
by WD following the square root law. 
The gravity darkening exponent $GR$ is between 0.2 and 0.6 in the
temperature range of 6300 and 8000K \citep[]{2017A&A...600A..30C,2021MNRAS.505.2051E}.
If we set $GR$ as a free parameter, the result 
converge at $\sim 0.3$. So we first adopt 0.3 for the current calculation and further extend to 0.5 to test
their effect on mass determination. 
Other parameters such as
the bolometric albedos $ALB$, 
the extinction in g,r and i band, and the distance ($\log pc$)
are listed in Table\ref{tab.wdfit}.
The phase folded ZTF light curves were then solved with Wilson-Devinney code.
The zero point fluxes ($CALIB$) for g, r, and i bands,
defined as 3631 Jy per unit frequency for all AB systems,
are chosen to be 0.49875, 0.288637 and 0.195711 $erg s^{-1} cm^{-3}$, respectively 
\footnote{http://svo2.cab.inta-csic.es/svo/theory/fps3/index.php}.

\begin{figure}[ht]
\center
\includegraphics[width=0.48\textwidth]{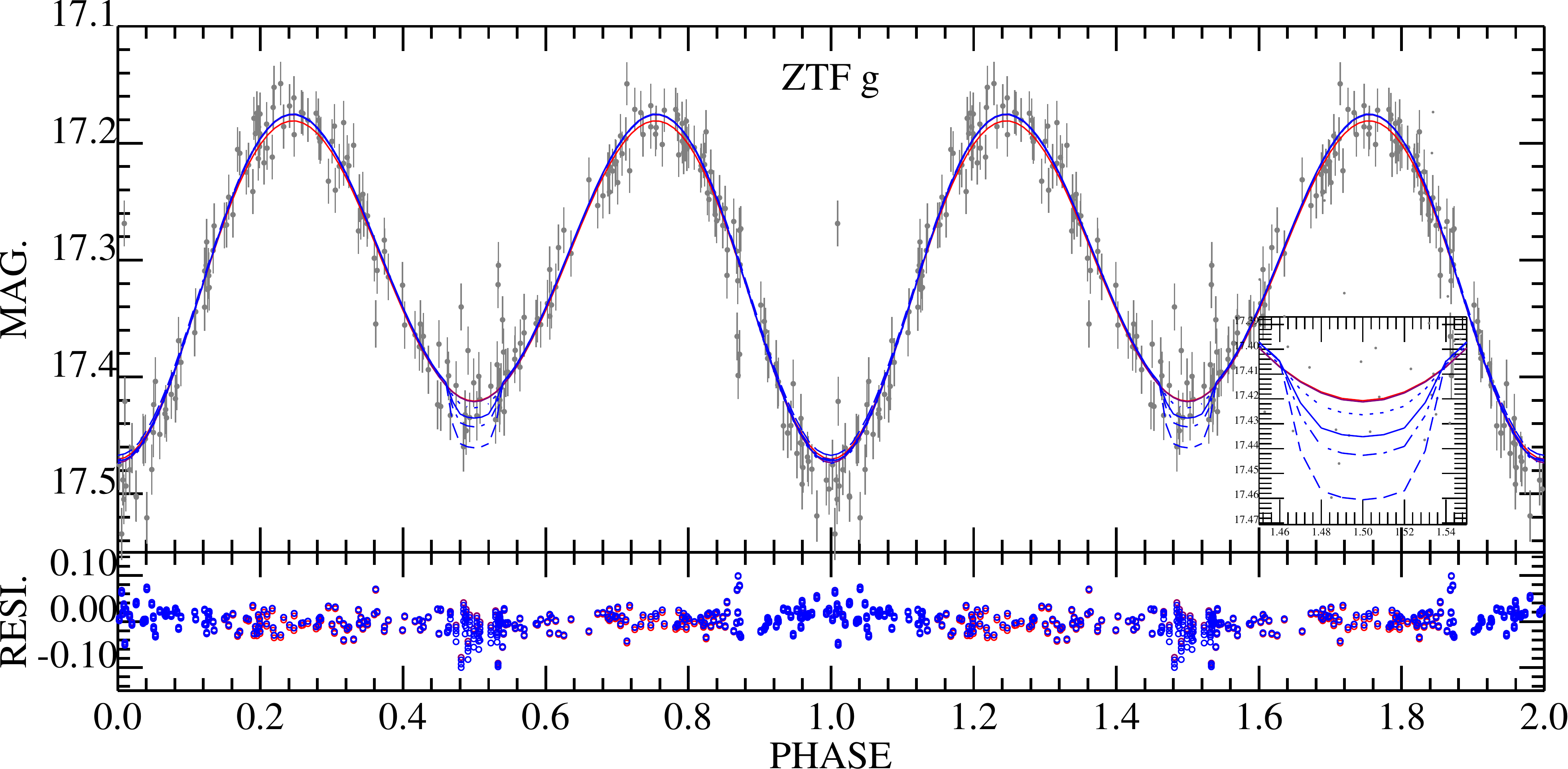}
\includegraphics[width=0.48\textwidth]{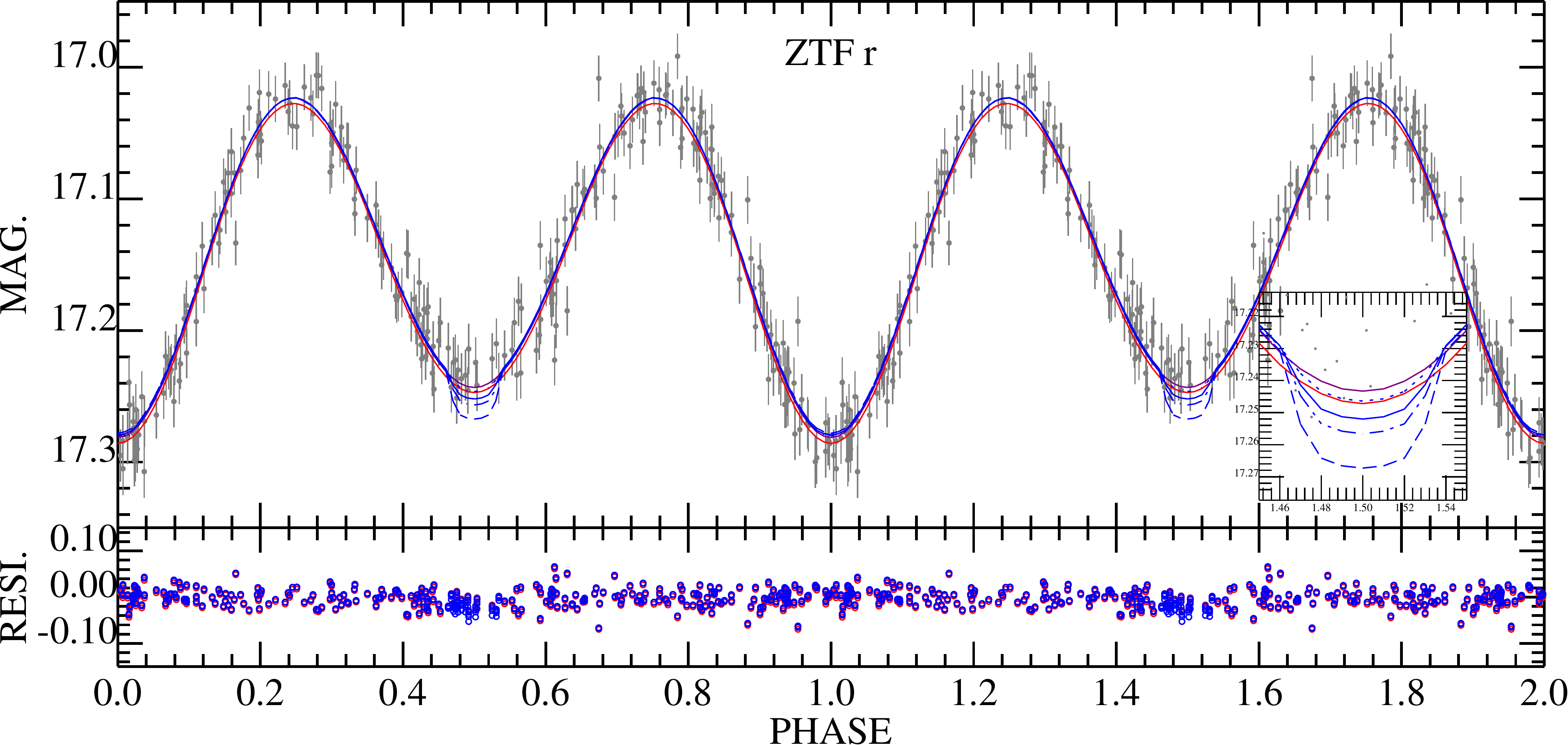}
\includegraphics[width=0.48\textwidth]{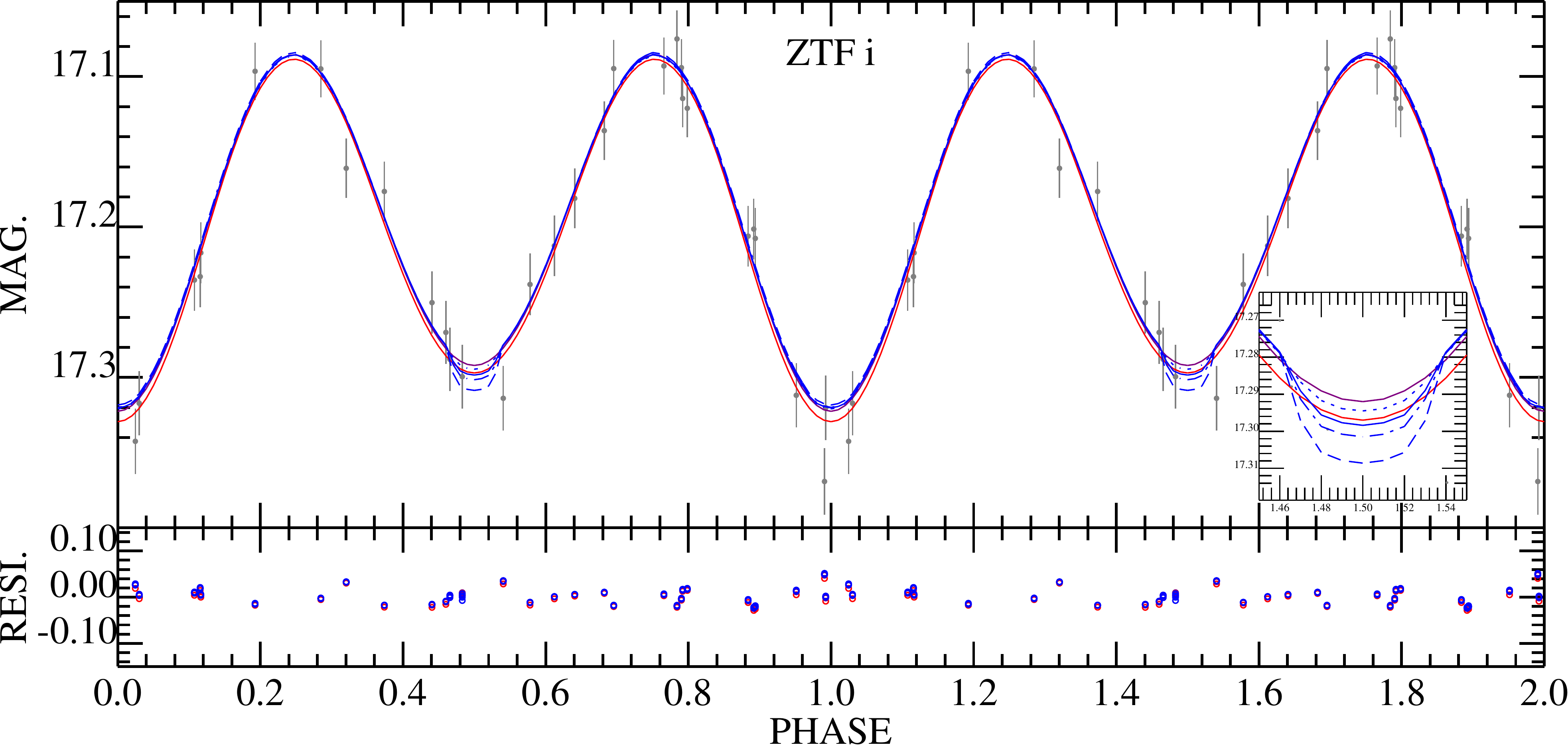}
\caption{Light curve fitting of ZTF g r and i band, 
from top to bottom, respectively.
The grey points in the upper panel denote ZTF magnitudes,
and the curves
are models calculated with Wilson-Devinney code,
with color from red to blue denote inclination increasing from $60^{\circ}$ to $90^{\circ}$. All the solid curve is fixed with white dwarf companion of 18000K. The fitting residuals are plotted in each lower panel, 
with a standard variation of $\approx2.7$\%.
To illustrate eclipse of white dwarf occultation in the light curve, models with
white dwarfs of $12000K, 22000K$ and $30000K$ are plotted with the dotted, the dot dash and the dashed line, respectively and enlarged in the inset of each plot, while the inclination is fixed at $90^{\circ}$.
} 
%
\label{fig.ztfgri_fit}
\end{figure}

The light curve fitting result is shown in Fig. \ref{fig.ztfgri_fit}.
The best fit model light curves with different inclination angles are illustrated.
In Fig. \ref{fig.ztfgri_fit}, the red color is for small inclination angle,
while the blue is for large inclination angle.
As the inclination angle changes, the mass ratio changes, then potential and orbital distance will make a compensatory change to fit the light curve. So in the range of inclination,
the difference in the light curve(0.1\%) is indistinguishable under the current photometric accuracy.
If the inclination angle is smaller ($\leq 60^{\circ}$),
the visible component will overfill its Roche lobe and transfer mass onto the compact star, 
and may possibly cause spectra emission features \citep{2021MNRAS.508.4106E}, which are however not seen in the spectra spanning 5 years.
Here we assume a detached configuration, 
but it is necessary to note that a low accretion rate scenario either do not display emission.
While for inclination larger than $80^\circ$, the occultation of star 2 will cause an eclipse in the light curve, and the depth of the eclipse depends on the temperature of the hot white dwarf, as illustrated in the inset of Fig. \ref{fig.ztfgri_fit}. 
As could be estimated from the mass function(i.e. equation\ref{eqmass}), when
the inclination is larger than $80^{\circ}$,
M2 varies from 0.84 to 1.00 $M_{\odot}$,
the corresponding radius is approximately 0.01 to 0.008$R_{\odot}$ \citep[][]{2005A&A...441..689A}.
Assuming a radius of $0.01R_{\odot}$ in the Wilson-Devinney code (specified by $POT2$),
the eclipse depth will be most prominent in g band, which 
is approximately 0.5\%, 1.5\%, 2.0\%, 4.0\% 
for T2 of 12000K, 18000K, 22000K and 30000K, respectively. 
When T2 is higher than 22000K, The eclipse will be deeper than 2\% in g band and will be detectable in current data. So if the inclination is higher than
$80^\circ$ and the companion is a white dwarf, the temperature of the white dwarf should be lower than 22000K, this is consistent with the estimation from SED fitting. 
As the eclipse is not detected, the model difference between different inclination is not distinguishable.
The upper limit of the inclination could not be determined.
Thus, we roughly limit the angle 
from 60$^{\circ}$ to 90$^{\circ}$.
Then a grid search upon inclination can be carried with Wilson-Devinney code.
The results show that
$SMA$ varies from 1.7 to 1.5 $R_{\odot}$,
$T1$ varies from 7400K to 7500K,
$q$ varies from 18 to 10,
and the filling factor (controlled by the star 1 potential $\Omega_{\rm 1}$) varies between 95\% and 100\%.
The corresponding mass of star 1 (the visible component) is from 0.07 to 0.09 $M_{\odot}$.

%

%
%

\begin{figure*}[htbp!]
\center
\includegraphics[width=0.96\textwidth]{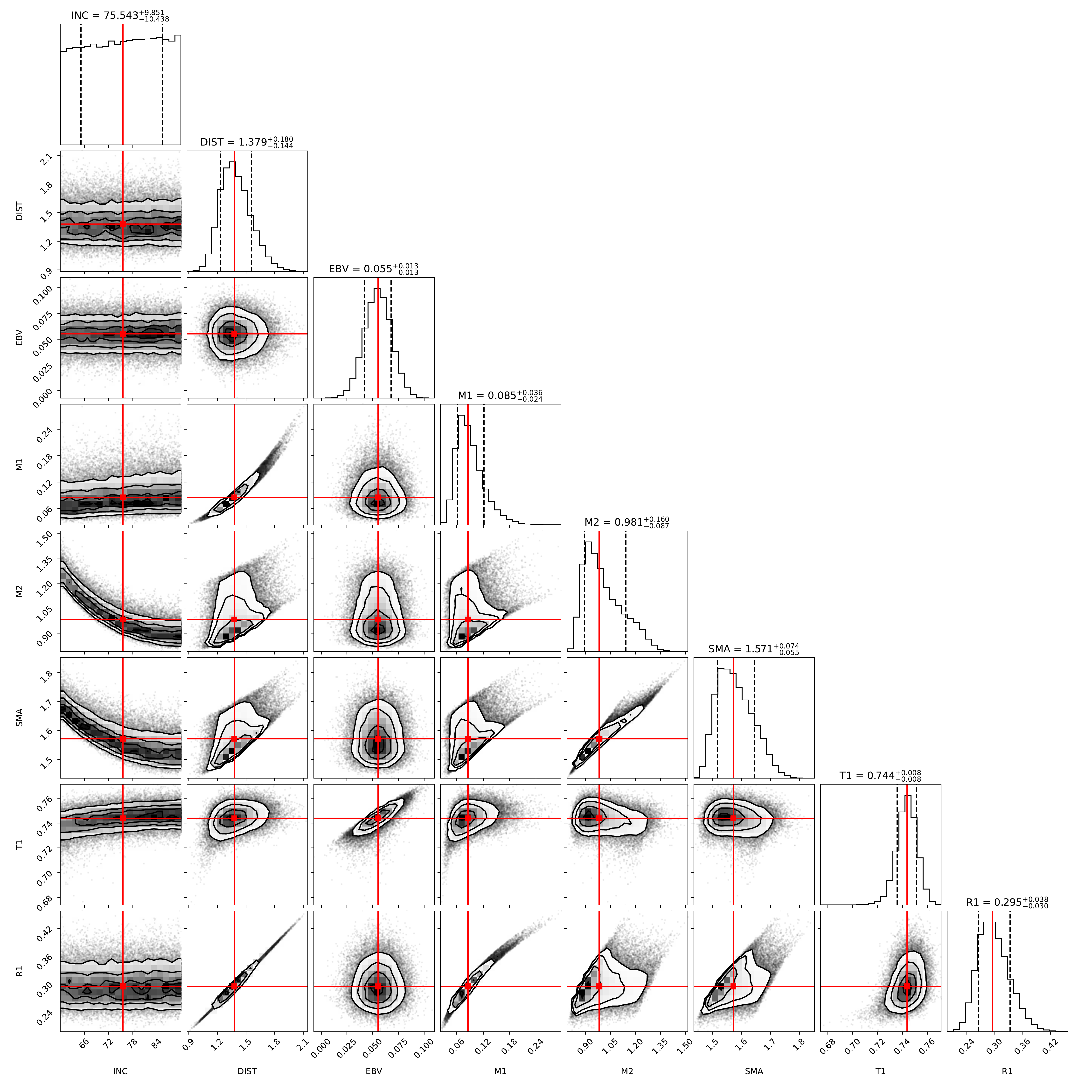}
\caption{
The bootstrapping sampling results upon ZTF g, r and i light curves, using Wilson-Devinney code.
Diagonal elements show the probability distribution of each parameter, with percentiles of 16\%, 50\% and 84\%, denoted by vertical lines.
Note star 1 denotes the visible component,
while star 2 denotes the invisible compact component.
The uncertainties of inclination, distance, extinction, gravity darkening and et al. are all taken into consideration.
See text for more details.
} 
\label{fig.ztfgri_mcmc}
\end{figure*}

To estimate the uncertainties of our solution,
we further repeat our process with larger range of distance, extinction,
gravity darkening and reflection coefficients,  and let the DC program of Wilson-Devinney code determine the $T1$, $q$, $SMA$, $INC$ and $POT1$,
via bootstrapping.
Inclination angles are chosen from 60$^{\circ}$ to 90$^{\circ}$ with probability distribution function (PDF) $=sin(i)$
(it is the equivalent to PDF($\cos i$)=1).
The distance is sampled according to Section \ref{sec:distance}.
Actually we do a Gaussian sampling ($mean=0.7248, \sigma=0.085375$) on parallax then get distance as $1/parallax$.
Extinction is sampled from Gaussian function with mean = 0.055 and sigma = 0.013.
Both the gravity darkening($GR$) and reflection coefficient($ALB$) are related to the status of the envelope of the star. For convective atmosphere $GR=0.3$ and $ALB=0.5$ is suggested, while for hot radiative atmosphere(e.g., main sequence star of $T_{eff} >8000K$), $1.0$ for both are suggested.
For our case, though the high temperature gradient in the bloated radius of star 1 will enable a convective envelope, but to be conservative, 
{$GR$ is sampled from uniform distribution within 0.3 to 0.5, while $ALB1$ is sampled from 0.5 to 0.7.}
For each set of parameters (distance, extinction, $GR$ and $ALB$) sampled,
we fix them in DC, 
change the light curve values by adding a Gaussian noise according to their errors,
run DC program of Wilson-Devinney code, 
and get optimized T1, SMA, Q and POT1({\bf Roche} potential). 
The related M1/2, R1, log $g1$ are also obtained.
We repeat the process by over 30000 times and calculate $\chi^2$, 
and the final results are shown in Fig. \ref{fig.ztfgri_mcmc},
giving 
$M1=0.085^{+0.036}_{-0.024}M_{\odot}$, R1=$0.295^{+0.038}_{-0.030}R_{\odot}$,
$SMA=1.57^{+0.07}_{-0.06}$, and $M2=0.98^{+0.16}_{-0.09}$.
The results correspond to star 1 log $g$ of $4.43\pm0.05$ and mass ratio Q of $12^{+5}_{-3}$.
The uncertainty of $M1$ mostly comes from the distance uncertainty. 
The gravity darkening efficient also has significant impact on the masses.
With stronger gravity darkening effect, the filling factor could be smaller, 
leading to larger orbital separation and hence larger masses for both component.

Within the bootstrapping sample, the $\chi^2-inclination$ relation is inspected and no correlation is found, 
confirming that the inclination cannot be constrained into narrower range unless additional constrains are considered.
The absence of emissions in the 5 year spectral observations, 
are considered as a sign of lacking ongoing mass transferring,
thus the system is modeled as a detached system.
The estimated filling factor is between 95\% and 100\%.
Such a hot but low mass visible component can only be properly explained as a pre-ELM, according to present day stellar evolution theories. 
Certainly the extremely low mass could be biased due to several reasons, like distance, gravity darkening effect, extinction and temperature.
Taking a step back, the estimation is consistent with the 0.14 $M_{\odot}$ theoretical limit under the 95\% confidence level. 
Follow up observations using larger aperture telescopes and higher precision instruments will be significant to refine the parameters.

\section{Summaries and Discussions}
\label{sec.sum}

\subsection{Formation of J2240}
\begin{figure}[htbp!]
\center
\includegraphics[width=0.48\textwidth]{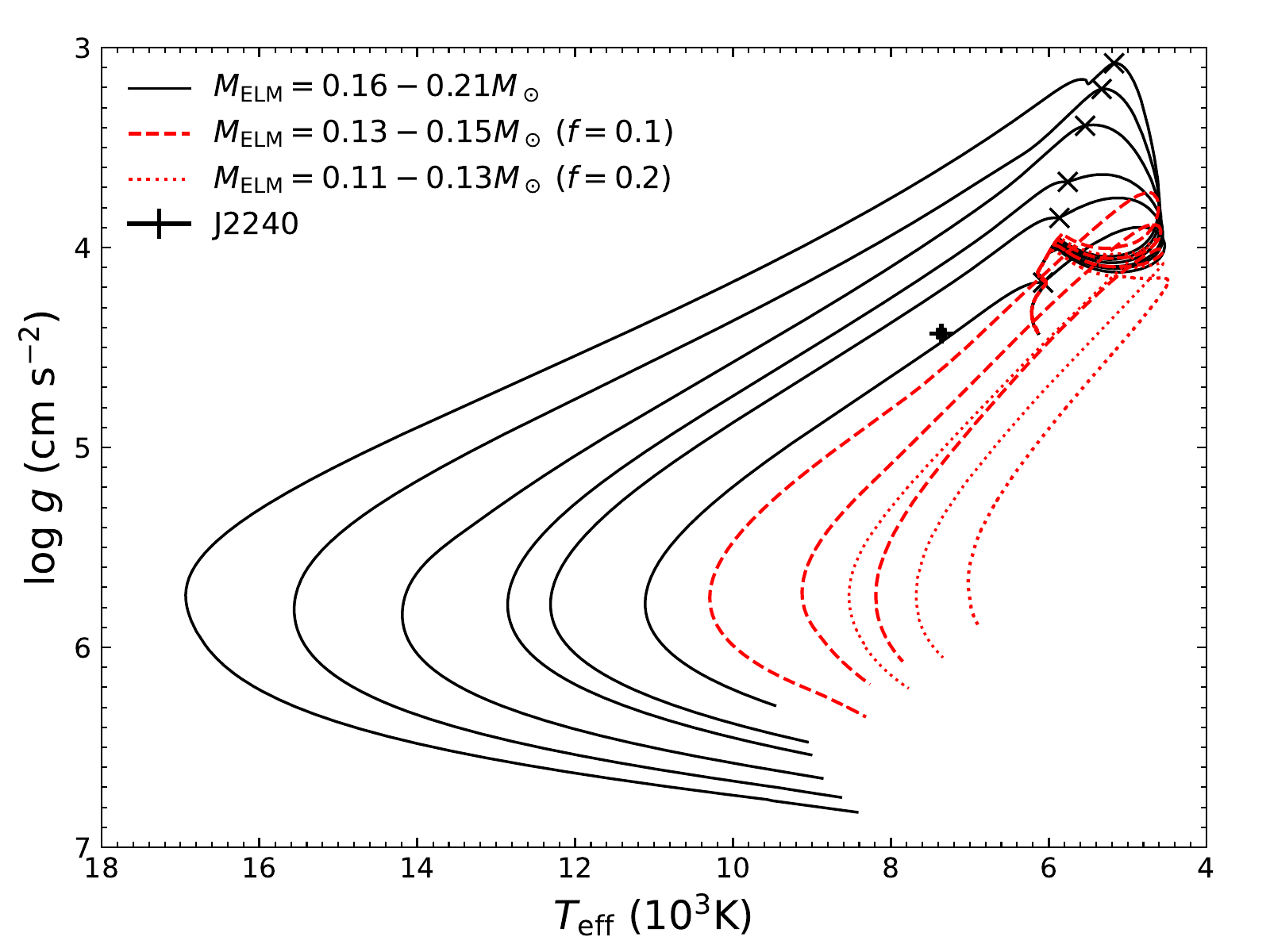}
\caption{The evolutionary tracks of ELM WDs in $\log\; g-\log\; T_{\rm eff}$ plane. 
Theses tracks have the same initial donor masses ($1.2M_{\odot}$) and accretor masses ($1.1M_{\odot}$). 
The standard evolutionary models of ELM WDs are shown in black solid lines with ELM masses of $0.16-0.21M_\odot$ in a step of $0.01M_\odot$ {\bf from right to left}. 
The cross symbols denote the end of mass transfer (MT) and the start of the contraction phase. 
Stars between the end of MT and the beginning of WD cooling track, generally referred as pre-ELMs, have nearly constant luminosity, 
decreasing radius and increasing temperature, and a lifetime of $\approx$ 1Gyr depending on their masses. 
The red lines are for the models with considering the pulsar evaporation effects. 
The models of $f=0.1$ are shown in the red dotted lines, where the ELMs have masses of $0.13,0.14,0.15M_\odot$ from right to left. 
The models of $f=0.2$ are shown in the red dashed lines, where the ELMs have masses of $0.11,0.12,0.13M_\odot$ from right to left.
{\bf J2240 here adopts $T_{\rm eff}=7360\pm150K$ and log $g=4.43\pm0.05$.}
}
\label{fig.elmtracks}
\end{figure}

\citet[][]{2019ApJ...871..148L} suggested that ELM WDs with mass less than $\sim 0.22M_\odot$ can not be produced from the CE ejection process due to the large binding energy of the progenitor envelope. Therefore, J2240 is supposed to be formed through stable Roche lobe overflow. 
To model the evolutionary history of J2240, we perform binary evolution simulations by using the stellar evolution code Modules for Experiments in Stellar Astrophysics (MESA, version r12115) \citep[][]{2011ApJS..192....3P, 2013ApJS..208....4P, 2015ApJS..220...15P, 2018ApJS..234...34P, 2019ApJS..243...10P}. The standard evolutionary tracks for ELM WDs from the stable Roche lobe overflow with mass from $0.16-0.21M_\odot$ (in steps of $0.01M_\odot$ from right to left) are presented in black lines of Fig. \ref{fig.elmtracks}, where the termination of mass transfer is marked with black crosses. After the mass transfer phase, the pre-ELM WDs are in the bloat state, and the radius gradually decreases.
The comparison between observation parameters of J2240 and the theoretical tracks suggests that J2240 may have a mass around $0.160-0.170M_\odot$. 
{\bf The estimation of ELM WD mass is only marginally consistent with the light curve fitting results within the 99\% confidence interval.}
Besides, the standard binary evolution models can not produce pre-ELM WDs with mass less than $\sim 0.14M_\odot$, 
as shown in previous works \citep[][]{2014A&A...571A..45I, 2017MNRAS.467.1874C, 2019ApJ...871..148L}. 

If J2240 {\bf has} a mass less than $\sim 0.14M_\odot$, the extra mass-loss mechanism for the progenitor of ELM WD is needed. For example, if the companion is a NS, the pulsar evaporation on the ELM WD progenitors can strip a part of envelope, which leads to low-mass remnants \citep[][]{2016ApJ...830..153J, 2021MNRAS.506.3323T}. Following \citet[][]{2016ApJ...830..153J}, we calculate several models with considering the pulsar evaporation effects. The corresponding mass-loss rate of the donor is given by 
\begin{equation}
\dot{M}_{\rm d,evap} = -\frac{f}{v^2_{\rm d,esc}}L_{\rm p}\left(\frac{R_{\rm d}}{a}\right),
\end{equation}
where $f$ is the efficiency of pulsar irradiation, $v_{\rm d,esc}$ and $R_{\rm d}$ are the surface escape velocity and the radius of the donor, respectively, $L_{\rm p}$ is the spin-down luminosity of the pulsar, and $a$ is the binary separation. Other initial parameters are similar as that in Model A of \citet[][]{2015ApJ...814...74J}.
The results of evaporation efficiency with $f=0.1,0.2$ are shown in Fig. \ref{fig.elmtracks}. We see that the ELM WDs with mass $\lesssim 0.14M_\odot$ can be produced, where the minimum ELM WD masses for $f=0.1,0.2$ are about $0.13M_\odot$ and $0.11M_\odot$, respectively. However, the evolutionary tracks for those ELM WDs with mass $\lesssim 0.14M_\odot$ are below the observation value of J2240. Therefore, the formation of J2240 gives a challenge to the standard evolutionary models of ELM WDs, and it deserves further detailed investigations \citep{2021ApJ...909..174D}. 

\subsection{Conclusion}

In this work, we present detailed mass estimation for ELM-WD J2240, 
which was discovered in the LAMOST low resolution survey as a binary system with the primary star orbiting an unseen compact star with a RV semi-amplitude of $K1=318.5\pm3.3km/s$ every 0.219658$\pm0.000002$ days. Kinematics analysis indicates that the system is a thin disk binary.
Accurate parallax from GAIA help to constraint the primary radius to $0.29\pm0.04R_{\odot}$. 
{\bf Temperatures from the spectral, SED and the multi-color light curve fitting are consistent with each other ($7360\pm150K$).}
The radial velocity curve combining with the ellipsoidal light curve offers a better constraint on the gravitational potential thus the mass providing that the temperature and the radius could be accurately determined. Considering the uncertainty in the input parameters, our best estimation on mass of the visible component is $M1=0.085^{+0.036}_{-0.024}M_{\odot}$, 
and the mass of unseen compact star is $M2=0.98^{+0.16}_{-0.09}M_{\odot}$.

Considering the mass and temperature of the primary star, it should be a helium white dwarf with very bloated hydrogen envelope, i.e. a pre-ELM WD, just finishing its mass transfer and entering the constant luminosity contracting phase. We note that the star was also listed as a pre-ELM WD candidate in table D1 of \citet{2021MNRAS.508.4106E} , but since there is no spectral observation, no mass estimation was given in that paper.
The mass of star 1 is relatively small compared to ELM theories, as shown in Fig. \ref{fig.elmtracks}.
Yet the small mass of visible component is also in accordance with 
the spectral estimated surface gravity (log $g$=4.14$\pm$0.25dex),
and the estimation from light curve fitting ($4.43\pm0.05dex$).
According to our current radius estimation,
to get a mass of $\approx0.14M_{\odot}$, a surface gravity of $\approx4.65$ is needed, which could be barely reached at the edge of our parameter estimation.
Since the visible component is almost filling its Roche lobe
(from $\approx95\%$ to $100\%$), 
the mass can be estimated from radius and the density period relation
\begin{equation}
\overline{\rho}_{donor} \approx 0.185 g{cm}^{-3}(P_{orb}/day)^{-2},
\end{equation}
with an accurate of 6\% \citep[][]{2022MNRAS.511L..24E}.
Using $R_1=0.29\pm0.03R_{\odot}$ and period of 0.219658 days,
the mass of star 1 is $0.07\pm0.02M_{\odot}$.
This can be used as an auxiliary verification.

The invisible companion has a mass of $0.98^{+0.16}_{-0.09}M_{\odot}$,
indicating a candidate of massive White Dwarf or a Neutron Star.
As no significant X-ray observation is found,
radio observation may be used to verify if it is a pulsar,
and deep ultraviolet spectral observation may be used to verify if it is a White Dwarf.
If it is a white dwarf, the temperature should be lower than 22000K based on the ultraviolet photometry of GALEX. 
A high cadence photometry with precision of 0.5\% will also be help to discern the shallow eclipse of a white dwarf.
Besides, the resolution and {\bf S/N} of current LAMOST and P200 spectra are too low to tell the chemical character of the primary.
High resolution spectral observations, using larger telescopes,
in the optical and near infrared bands, may reveal clues on the evolution history of J2240.
The discovery and confirmation of more binaries hosting compact components using time domain spectral survey \citep{2022SCPMA..6529711M, 2022ApJ...938...78L, 2022ApJ...933..193Z, 2022ApJ...940..165Y}
is important in improving the understanding of stellar evolution theories.


\section*{acknowledgements}
Guoshoujing Telescope (the Large Sky Area Multi-Object Fiber Spectroscopic Telescope LAMOST) is a National Major Scientific Project built by the Chinese Academy of Sciences. Funding for the project has been provided by the National Development and Reform Commission. LAMOST is operated and managed by the National Astronomical Observatories, Chinese Academy of Sciences. 
This work presents results from the European Space Agency (ESA) space mission Gaia. Gaia data are being processed by the Gaia Data Processing and Analysis Consortium (DPAC). Funding for the DPAC is provided by national institutions, in particular the institutions participating in the Gaia MultiLateral Agreement (MLA). The Gaia mission website is https://www.cosmos.esa.int/gaia. The Gaia archive website is https://archives.esac.esa.int/gaia. We acknowledge use of the VizieR catalogue access tool, operated at CDS, Strasbourg, France, and of Astropy, a community-developed core Python package for Astronomy (Astropy Collaboration, 2013). 
Z.H.T. would like to thank Fu Xiaoting, Zhang Xiaobin, Shao Yong, Liu Jifeng an Wangsong for the useful discussions and suggestions.
Z.H.T. thanks the support of the National Key R\&D Program of China (2019YFA0405000), NSFC 12090041 and NSFC 11933004. 
Y.H.L. acknowledges support from the Youth Innovation Promotion Association of the CAS (Id. 2020060) and National Natural Science Foundation of China (Grant No. 11873066). 
ZWL and XFC thanks the support of National Key R\&D Program of China (Gant No. 2021YFA1600403), 
National Natural Science Foundation of China (Grant Nos. 11733008, 12103086, 12090040/12090043),
the National Science Fund for Distinguished Young Scholars (Grant No. 12125303)
and the Yunnan Fundamental Research Projects (No. 202101AU070276). 
We also acknowledge the science research grant from the China Manned Space Project with No.CMS-CSST-2021-A10.

\bibliographystyle{aasjournal}
\bibliography{bibtex.bib}{}

\begin{thebibliography}{}
\expandafter\ifx\csname natexlab\endcsname\relax\def\natexlab#1{#1}\fi
\providecommand{\url}[1]{\href{#1}{#1}}
\providecommand{\dodoi}[1]{doi:~\href{http://doi.org/#1}{\nolinkurl{#1}}}
\providecommand{\doeprint}[1]{\href{http://ascl.net/#1}{\nolinkurl{http://ascl.net/#1}}}
\providecommand{\doarXiv}[1]{\href{https://arxiv.org/abs/#1}{\nolinkurl{https://arxiv.org/abs/#1}}}

\bibitem[{{Ahumada} {et~al.}(2020){Ahumada}, {Prieto}, {Almeida}, {Anders},
  {Anderson}, {Andrews}, {Anguiano}, {Arcodia}, {Armengaud}, {Aubert}, \&
  et~al.}]{2020ApJS..249....3A}
{Ahumada}, R., {Prieto}, C.~A., {Almeida}, A., {et~al.} 2020, \apjs, 249, 3,
  \dodoi{10.3847/1538-4365/ab929e}

\bibitem[{{Althaus} {et~al.}(2005){Althaus}, {Garc{\'\i}a-Berro}, {Isern}, \&
  {C{\'o}rsico}}]{2005A&A...441..689A}
{Althaus}, L.~G., {Garc{\'\i}a-Berro}, E., {Isern}, J., \& {C{\'o}rsico}, A.~H.
  2005, \aap, 441, 689, \dodoi{10.1051/0004-6361:20052996}

\bibitem[{{Bai} {et~al.}(2021){Bai}, {Zhang}, {Yuan}, {Fan}, {He}, {Lei},
  {Dong}, {Yu}, {Zhao}, {Zhang}, {Hou}, \& {Chu}}]{2021RAA....21..249B}
{Bai}, Z.-R., {Zhang}, H.-T., {Yuan}, H.-L., {et~al.} 2021, Research in
  Astronomy and Astrophysics, 21, 249, \dodoi{10.1088/1674-4527/21/10/249}

\bibitem[{{Bailer-Jones} {et~al.}(2021){Bailer-Jones}, {Rybizki}, {Fouesneau},
  {Demleitner}, \& {Andrae}}]{2021AJ....161..147B}
{Bailer-Jones}, C.~A.~L., {Rybizki}, J., {Fouesneau}, M., {Demleitner}, M., \&
  {Andrae}, R. 2021, \aj, 161, 147, \dodoi{10.3847/1538-3881/abd806}

\bibitem[{{Bellm} {et~al.}(2019){Bellm}, {Kulkarni}, {Graham}, {Dekany}, \&
  {Smith}}]{2019pasp..131.018002}
{Bellm}, E., {Kulkarni}, S., {Graham}, M., {Dekany}, R., \& {Smith}, R. 2019,
  \pasp, 131, 018002, \dodoi{10.1088/1538-3873/aaecbe}

\bibitem[{{Bianchi} {et~al.}(2017){Bianchi}, {Shiao}, \&
  {Thilker}}]{2017ApJS..230...24B}
{Bianchi}, L., {Shiao}, B., \& {Thilker}, D. 2017, \apjs, 230, 24,
  \dodoi{10.3847/1538-4365/aa7053}

\bibitem[{{Brown} {et~al.}(2016){Brown}, {Gianninas}, {Kilic}, {Kenyon}, \&
  {Allende Prieto}}]{2016ApJ...818..155B}
{Brown}, W.~R., {Gianninas}, A., {Kilic}, M., {Kenyon}, S.~J., \& {Allende
  Prieto}, C. 2016, \apj, 818, 155, \dodoi{10.3847/0004-637X/818/2/155}

\bibitem[{{Brown} {et~al.}(2010){Brown}, {Kilic}, {Allende Prieto}, \&
  {Kenyon}}]{2010ApJ...723.1072B}
{Brown}, W.~R., {Kilic}, M., {Allende Prieto}, C., \& {Kenyon}, S.~J. 2010,
  \apj, 723, 1072, \dodoi{10.1088/0004-637X/723/2/1072}

\bibitem[{{Brown} {et~al.}(2017){Brown}, {Kilic}, \&
  {Gianninas}}]{2017ApJ...839...23B}
{Brown}, W.~R., {Kilic}, M., \& {Gianninas}, A. 2017, \apj, 839, 23,
  \dodoi{10.3847/1538-4357/aa67e4}

\bibitem[{{Cardelli} {et~al.}(1989){Cardelli}, {Clayton}, \&
  {Mathis}}]{1989ApJ...345..245C}
{Cardelli}, J.~A., {Clayton}, G.~C., \& {Mathis}, J.~S. 1989, \apj, 345, 245,
  \dodoi{10.1086/167900}

\bibitem[{{Castelli} \& {Kurucz}(2003)}]{Castelli2003}
{Castelli}, F., \& {Kurucz}, R.~L. 2003, in Modelling of Stellar Atmospheres,
  ed. N.~{Piskunov}, W.~W. {Weiss}, \& D.~F. {Gray}, Vol. 210, A20.
\newblock \doarXiv{astro-ph/0405087}

\bibitem[{{Chambers} {et~al.}(2016){Chambers}, {Magnier}, {Metcalfe},
  {Flewelling}, {Huber}, {Waters}, {Denneau}, {Draper}, {Farrow}, {Finkbeiner},
  {Holmberg}, {Koppenhoefer}, {Price}, {Rest}, {Saglia}, {Schlafly}, {Smartt},
  {Sweeney}, {Wainscoat}, {Burgett}, {Chastel}, {Grav}, {Heasley}, {Hodapp},
  {Jedicke}, {Kaiser}, {Kudritzki}, {Luppino}, {Lupton}, {Monet}, {Morgan},
  {Onaka}, {Shiao}, {Stubbs}, {Tonry}, {White}, {Ba{\~n}ados}, {Bell},
  {Bender}, {Bernard}, {Boegner}, {Boffi}, {Botticella}, {Calamida},
  {Casertano}, {Chen}, {Chen}, {Cole}, {Deacon}, {Frenk}, {Fitzsimmons},
  {Gezari}, {Gibbs}, {Goessl}, {Goggia}, {Gourgue}, {Goldman}, {Grant},
  {Grebel}, {Hambly}, {Hasinger}, {Heavens}, {Heckman}, {Henderson}, {Henning},
  {Holman}, {Hopp}, {Ip}, {Isani}, {Jackson}, {Keyes}, {Koekemoer}, {Kotak},
  {Le}, {Liska}, {Long}, {Lucey}, {Liu}, {Martin}, {Masci}, {McLean}, {Mindel},
  {Misra}, {Morganson}, {Murphy}, {Obaika}, {Narayan}, {Nieto-Santisteban},
  {Norberg}, {Peacock}, {Pier}, {Postman}, {Primak}, {Rae}, {Rai}, {Riess},
  {Riffeser}, {Rix}, {R{\"o}ser}, {Russel}, {Rutz}, {Schilbach}, {Schultz},
  {Scolnic}, {Strolger}, {Szalay}, {Seitz}, {Small}, {Smith}, {Soderblom},
  {Taylor}, {Thomson}, {Taylor}, {Thakar}, {Thiel}, {Thilker}, {Unger},
  {Urata}, {Valenti}, {Wagner}, {Walder}, {Walter}, {Watters}, {Werner},
  {Wood-Vasey}, \& {Wyse}}]{2016arXiv161205560C}
{Chambers}, K.~C., {Magnier}, E.~A., {Metcalfe}, N., {et~al.} 2016, arXiv
  e-prints, arXiv:1612.05560.
\newblock \doarXiv{1612.05560}

\bibitem[{{Chen} {et~al.}(2017){Chen}, {Maxted}, {Li}, \&
  {Han}}]{2017MNRAS.467.1874C}
{Chen}, X., {Maxted}, P.~F.~L., {Li}, J., \& {Han}, Z. 2017, \mnras, 467, 1874,
  \dodoi{10.1093/mnras/stx115}

\bibitem[{{Chen} {et~al.}(2020){Chen}, {Wang}, {Deng}, {de Grijs}, {Yang}, \&
  {Tian}}]{2020ApJS..249...18C}
{Chen}, X., {Wang}, S., {Deng}, L., {et~al.} 2020, \apjs, 249, 18,
  \dodoi{10.3847/1538-4365/ab9cae}

\bibitem[{{Claret}(2017)}]{2017A&A...600A..30C}
{Claret}, A. 2017, \aap, 600, A30, \dodoi{10.1051/0004-6361/201629705}

\bibitem[{{Cui} {et~al.}(2012){Cui}, {Zhao}, {Chu}, {Li}, {Li}, {Zhang}, {Su},
  {Yao}, {Wang}, {Xing}, {Li}, {Zhu}, {Wang}, {Gu}, {Luo}, {Xu}, {Zhang},
  {Liu}, {Zhang}, {Yang}, {Cao}, {Chen}, {Chen}, {Chen}, {Chen}, {Chu}, {Feng},
  {Gong}, {Hou}, {Hu}, {Hu}, {Hu}, {Jia}, {Jiang}, {Jiang}, {Jiang}, {Jin},
  {Li}, {Li}, {Li}, {Liu}, {Liu}, {Lu}, {Mao}, {Men}, {Qi}, {Qi}, {Shi},
  {Tang}, {Tao}, {Wang}, {Wang}, {Wang}, {Wang}, {Wang}, {Wang}, {Wang},
  {Wang}, {Wang}, {Wang}, {Wang}, {Wang}, {Xu}, {Xu}, {Yang}, {Yu}, {Yuan},
  {Yuan}, {Zhai}, {Zhang}, {Zhang}, {Zhang}, {Zhao}, {Zhou}, {Zhou}, {Zhu}, \&
  {Zou}}]{2012RAA....12.1197C}
{Cui}, X.-Q., {Zhao}, Y.-H., {Chu}, Y.-Q., {et~al.} 2012, Research in Astronomy
  and Astrophysics, 12, 1197, \dodoi{10.1088/1674-4527/12/9/003}

\bibitem[{{Cutri} {et~al.}(2021){Cutri}, {Wright}, {Conrow}, {Fowler},
  {Eisenhardt}, {Grillmair}, {Kirkpatrick}, {Masci}, {McCallon}, {Wheelock},
  {Fajardo-Acosta}, {Yan}, {Benford}, {Harbut}, {Jarrett}, {Lake}, {Leisawitz},
  {Ressler}, {Stanford}, {Tsai}, {Liu}, {Helou}, {Mainzer}, {Gettngs},
  {Gonzalez}, {Hoffman}, {Marsh}, {Padgett}, {Skrutskie}, {Beck}, {Papin}, \&
  {Wittman}}]{2014yCat.2328....0C}
{Cutri}, R.~M., {Wright}, E.~L., {Conrow}, T., {et~al.} 2021, VizieR Online
  Data Catalog, II/328

\bibitem[{{Deng} {et~al.}(2021){Deng}, {Li}, {Gao}, \&
  {Shao}}]{2021ApJ...909..174D}
{Deng}, Z.-L., {Li}, X.-D., {Gao}, Z.-F., \& {Shao}, Y. 2021, \apj, 909, 174,
  \dodoi{10.3847/1538-4357/abe0b2}

\bibitem[{{Drake} {et~al.}(2014){Drake}, {Graham}, {Djorgovski}, {Catelan},
  {Mahabal}, {Torrealba}, {Garc{\'\i}a-{\'A}lvarez}, {Donalek}, {Prieto},
  {Williams}, {Larson}, {Christen sen}, {Belokurov}, {Koposov}, {Beshore},
  {Boattini}, {Gibbs}, {Hill}, {Kowalski}, {Johnson}, \&
  {Shelly}}]{2014ApJS..213....9D}
{Drake}, A.~J., {Graham}, M.~J., {Djorgovski}, S.~G., {et~al.} 2014, \apjs,
  213, 9, \dodoi{10.1088/0067-0049/213/1/9}

\bibitem[{{El-Badry} \& {Burdge}(2022)}]{2022MNRAS.511L..24E}
{El-Badry}, K., \& {Burdge}, K.~B. 2022, \mnras, 511, 24,
  \dodoi{10.1093/mnrasl/slab135}

\bibitem[{{El-Badry} {et~al.}(2021{\natexlab{a}}){El-Badry}, {Rix}, {Quataert},
  {Kupfer}, \& {Shen}}]{2021MNRAS.508.4106E}
{El-Badry}, K., {Rix}, H.-W., {Quataert}, E., {Kupfer}, T., \& {Shen}, K.~J.
  2021{\natexlab{a}}, \mnras, 508, 4106, \dodoi{10.1093/mnras/stab2583}

\bibitem[{{El-Badry} {et~al.}(2021{\natexlab{b}}){El-Badry}, {Quataert}, {Rix},
  {Weisz}, {Kupfer}, {Shen}, {Xiang}, {Yang}, \& {Liu}}]{2021MNRAS.505.2051E}
{El-Badry}, K., {Quataert}, E., {Rix}, H.-W., {et~al.} 2021{\natexlab{b}},
  \mnras, 505, 2051, \dodoi{10.1093/mnras/stab1318}

\bibitem[{{Gaia Collaboration} {et~al.}(2021){Gaia Collaboration}, {Brown},
  {Vallenari}, {Prusti}, {de Bruijne}, {Babusiaux}, {Biermann}, {Creevey},
  {Evans}, {Eyer}, {Hutton}, {Jansen}, {Jordi}, {Klioner}, {Lammers},
  {Lindegren}, {Luri}, {Mignard}, {Panem}, {Pourbaix}, {Randich}, {Sartoretti},
  {Soubiran}, {Walton}, {Arenou}, {Bailer-Jones}, {Bastian}, {Cropper},
  {Drimmel}, {Katz}, {Lattanzi}, {van Leeuwen}, {Bakker}, {Cacciari},
  {Casta{\~n}eda}, {De Angeli}, {Ducourant}, {Fabricius}, {Fouesneau},
  {Fr{\'e}mat}, {Guerra}, {Guerrier}, {Guiraud}, {Jean-Antoine Piccolo},
  {Masana}, {Messineo}, {Mowlavi}, {Nicolas}, {Nienartowicz}, {Pailler},
  {Panuzzo}, {Riclet}, {Roux}, {Seabroke}, {Sordo}, {Tanga}, {Th{\'e}venin},
  {Gracia-Abril}, {Portell}, {Teyssier}, {Altmann}, {Andrae}, {Bellas-Velidis},
  {Benson}, {Berthier}, {Blomme}, {Brugaletta}, {Burgess}, {Busso}, {Carry},
  {Cellino}, {Cheek}, {Clementini}, {Damerdji}, {Davidson}, {Delchambre},
  {Dell'Oro}, {Fern{\'a}ndez-Hern{\'a}ndez}, {Galluccio}, {Garc{\'\i}a-Lario},
  {Garcia-Reinaldos}, {Gonz{\'a}lez-N{\'u}{\~n}ez}, {Gosset}, {Haigron},
  {Halbwachs}, {Hambly}, {Harrison}, {Hatzidimitriou}, {Heiter},
  {Hern{\'a}ndez}, {Hestroffer}, {Hodgkin}, {Holl}, {Jan{\ss}en}, {Jevardat de
  Fombelle}, {Jordan}, {Krone-Martins}, {Lanzafame}, {L{\"o}ffler}, {Lorca},
  {Manteiga}, {Marchal}, {Marrese}, {Moitinho}, {Mora}, {Muinonen}, {Osborne},
  {Pancino}, {Pauwels}, {Petit}, {Recio-Blanco}, {Richards}, {Riello},
  {Rimoldini}, {Robin}, {Roegiers}, {Rybizki}, {Sarro}, {Siopis}, {Smith},
  {Sozzetti}, {Ulla}, {Utrilla}, {van Leeuwen}, {van Reeven}, {Abbas}, {Abreu
  Aramburu}, {Accart}, {Aerts}, {Aguado}, {Ajaj}, {Altavilla}, {{\'A}lvarez},
  {{\'A}lvarez Cid-Fuentes}, {Alves}, {Anderson}, {Anglada Varela}, {Antoja},
  {Audard}, {Baines}, {Baker}, {Balaguer-N{\'u}{\~n}ez}, {Balbinot}, {Balog},
  {Barache}, {Barbato}, {Barros}, {Barstow}, {Bartolom{\'e}}, {Bassilana},
  {Bauchet}, {Baudesson-Stella}, {Becciani}, {Bellazzini}, {Bernet}, {Bertone},
  {Bianchi}, {Blanco-Cuaresma}, {Boch}, {Bombrun}, {Bossini}, {Bouquillon},
  {Bragaglia}, {Bramante}, {Breedt}, {Bressan}, {Brouillet}, {Bucciarelli},
  {Burlacu}, {Busonero}, {Butkevich}, {Buzzi}, {Caffau}, {Cancelliere},
  {C{\'a}novas}, {Cantat-Gaudin}, {Carballo}, {Carlucci}, {Carnerero},
  {Carrasco}, {Casamiquela}, {Castellani}, {Castro-Ginard}, {Castro Sampol},
  {Chaoul}, {Charlot}, {Chemin}, {Chiavassa}, {Cioni}, {Comoretto}, {Cooper},
  {Cornez}, {Cowell}, {Crifo}, {Crosta}, {Crowley}, {Dafonte}, {Dapergolas},
  {David}, {David}, {de Laverny}, {De Luise}, {De March}, {De Ridder}, {de
  Souza}, {de Teodoro}, {de Torres}, {del Peloso}, {del Pozo}, {Delbo},
  {Delgado}, {Delgado}, {Delisle}, {Di Matteo}, {Diakite}, {Diener},
  {Distefano}, {Dolding}, {Eappachen}, {Edvardsson}, {Enke}, {Esquej}, {Fabre},
  {Fabrizio}, {Faigler}, {Fedorets}, {Fernique}, {Fienga}, {Figueras},
  {Fouron}, {Fragkoudi}, {Fraile}, {Franke}, {Gai}, {Garabato},
  {Garcia-Gutierrez}, {Garc{\'\i}a-Torres}, {Garofalo}, {Gavras}, {Gerlach},
  {Geyer}, {Giacobbe}, {Gilmore}, {Girona}, {Giuffrida}, {Gomel}, {Gomez},
  {Gonzalez-Santamaria}, {Gonz{\'a}lez-Vidal}, {Granvik},
  {Guti{\'e}rrez-S{\'a}nchez}, {Guy}, {Hauser}, {Haywood}, {Helmi}, {Hidalgo},
  {Hilger}, {H{\l}adczuk}, {Hobbs}, {Holland}, {Huckle}, {Jasniewicz},
  {Jonker}, {Juaristi Campillo}, {Julbe}, {Karbevska}, {Kervella}, {Khanna},
  {Kochoska}, {Kontizas}, {Kordopatis}, {Korn}, {Kostrzewa-Rutkowska},
  {Kruszy{\'n}ska}, {Lambert}, {Lanza}, {Lasne}, {Le Campion}, {Le Fustec},
  {Lebreton}, {Lebzelter}, {Leccia}, {Leclerc}, {Lecoeur-Taibi}, {Liao},
  {Licata}, {Lindstr{\o}m}, {Lister}, {Livanou}, {Lobel}, {Madrero Pardo},
  {Managau}, {Mann}, {Marchant}, {Marconi}, {Marcos Santos}, {Marinoni},
  {Marocco}, {Marshall}, {Martin Polo}, {Mart{\'\i}n-Fleitas}, {Masip},
  {Massari}, {Mastrobuono-Battisti}, {Mazeh}, {McMillan}, {Messina},
  {Michalik}, {Millar}, {Mints}, {Molina}, {Molinaro}, {Moln{\'a}r},
  {Montegriffo}, {Mor}, {Morbidelli}, {Morel}, {Morris}, {Mulone}, {Munoz},
  {Muraveva}, {Murphy}, {Musella}, {Noval}, {Ord{\'e}novic}, {Orr{\`u}},
  {Osinde}, {Pagani}, {Pagano}, {Palaversa}, {Palicio}, {Panahi}, {Pawlak},
  {Pe{\~n}alosa Esteller}, {Penttil{\"a}}, {Piersimoni}, {Pineau}, {Plachy},
  {Plum}, {Poggio}, {Poretti}, {Poujoulet}, {Pr{\v{s}}a}, {Pulone}, {Racero},
  {Ragaini}, {Rainer}, {Raiteri}, {Rambaux}, {Ramos}, {Ramos-Lerate}, {Re
  Fiorentin}, {Regibo}, {Reyl{\'e}}, {Ripepi}, {Riva}, {Rixon}, {Robichon},
  {Robin}, {Roelens}, {Rohrbasser}, {Romero-G{\'o}mez}, {Rowell}, {Royer},
  {Rybicki}, {Sadowski}, {Sagrist{\`a} Sell{\'e}s}, {Sahlmann}, {Salgado},
  {Salguero}, {Samaras}, {Sanchez Gimenez}, {Sanna}, {Santove{\~n}a},
  {Sarasso}, {Schultheis}, {Sciacca}, {Segol}, {Segovia}, {S{\'e}gransan},
  {Semeux}, {Shahaf}, {Siddiqui}, {Siebert}, {Siltala}, {Slezak}, {Smart},
  {Solano}, {Solitro}, {Souami}, {Souchay}, {Spagna}, {Spoto}, {Steele},
  {Steidelm{\"u}ller}, {Stephenson}, {S{\"u}veges}, {Szabados}, {Szegedi-Elek},
  {Taris}, {Tauran}, {Taylor}, {Teixeira}, {Thuillot}, {Tonello}, {Torra},
  {Torra}, {Turon}, {Unger}, {Vaillant}, {van Dillen}, {Vanel}, {Vecchiato},
  {Viala}, {Vicente}, {Voutsinas}, {Weiler}, {Wevers}, {Wyrzykowski}, {Yoldas},
  {Yvard}, {Zhao}, {Zorec}, {Zucker}, {Zurbach}, \& {Zwitter}}]{2021Gaia}
{Gaia Collaboration}, {Brown}, A.~G.~A., {Vallenari}, A., {et~al.} 2021, \aap,
  649, A1, \dodoi{10.1051/0004-6361/202039657}

\bibitem[{{Gomel} {et~al.}(2021){Gomel}, {Faigler}, \&
  {Mazeh}}]{2021MNRAS.501.2822G}
{Gomel}, R., {Faigler}, S., \& {Mazeh}, T. 2021, \mnras, 501, 2822,
  \dodoi{10.1093/mnras/staa3305}

\bibitem[{{Green} {et~al.}(2019){Green}, {Schlafly}, {Zucker}, {Speagle}, \&
  {Finkbeiner}}]{2019ApJ...887...93G}
{Green}, G.~M., {Schlafly}, E., {Zucker}, C., {Speagle}, J.~S., \&
  {Finkbeiner}, D. 2019, \apj, 887, 93, \dodoi{10.3847/1538-4357/ab5362}

\bibitem[{{Han} {et~al.}(2002){Han}, {Podsiadlowski}, {Maxted}, {Marsh}, \&
  {Ivanova}}]{2002MNRAS.336..449H}
{Han}, Z., {Podsiadlowski}, P., {Maxted}, P.~F.~L., {Marsh}, T.~R., \&
  {Ivanova}, N. 2002, \mnras, 336, 449,
  \dodoi{10.1046/j.1365-8711.2002.05752.x}

\bibitem[{{He} {et~al.}(2019){He}, {Wang}, {Xu}, {Soria}, {Liu}, {Li}, {Bai},
  {Bai}, {Guo}, {Qiu}, {Zhang}, {Xu}, \& {Qian}}]{2019RAA....19...98H}
{He}, L., {Wang}, S., {Xu}, X.-J., {et~al.} 2019, Research in Astronomy and
  Astrophysics, 19, 098, \dodoi{10.1088/1674-4527/19/7/98}

\bibitem[{{Heinze} {et~al.}(2018){Heinze}, {Tonry}, {Denneau}, {Flewelling},
  {Stalder}, {Rest}, {Smith}, {Smartt}, \& {Weiland}}]{2018AJ....156..241H}
{Heinze}, A.~N., {Tonry}, J.~L., {Denneau}, L., {et~al.} 2018, \aj, 156, 241,
  \dodoi{10.3847/1538-3881/aae47f}

\bibitem[{{Hermes} {et~al.}(2012){Hermes}, {Kilic}, {Brown}, {Montgomery}, \&
  {Winget}}]{2012ApJ..74942}
{Hermes}, J., {Kilic}, M., {Brown}, W., {Montgomery}, M., \& {Winget}, D. 2012,
  \apj, 749, 42, \dodoi{10.1088/0004-637X/749/1/42}

\bibitem[{{Iben} \& {Tutukov}(1986)}]{1986ApJ...311..742I}
{Iben}, Icko, J., \& {Tutukov}, A.~V. 1986, \apj, 311, 742,
  \dodoi{10.1086/164812}

\bibitem[{{Istrate} {et~al.}(2016){Istrate}, {Marchant}, {Tauris}, {Langer}, \&
  {Stamcliffe}}]{2016A&A...595...35}
{Istrate}, A.~G., {Marchant}, P., {Tauris}, T.~M., {Langer}, N., \&
  {Stamcliffe}, R. 2016, \aap, 595, A35, \dodoi{10.1051/0004-6361/201628874}

\bibitem[{{Istrate} {et~al.}(2014){Istrate}, {Tauris}, \&
  {Langer}}]{2014A&A...571A..45I}
{Istrate}, A.~G., {Tauris}, T.~M., \& {Langer}, N. 2014, \aap, 571, A45,
  \dodoi{10.1051/0004-6361/201424680}

\bibitem[{{Jayasinghe} {et~al.}(2019){Jayasinghe}, {Stanek}, {Kochanek},
  {Shappee}, {Holoien}, {Thompson}, {Prieto}, {Dong}, {Pawlak}, {Pejcha},
  {Shields}, {Pojmanski}, {Otero}, {Britt}, \& {Will}}]{2019MNRAS.486.1907J}
{Jayasinghe}, T., {Stanek}, K.~Z., {Kochanek}, C.~S., {et~al.} 2019, \mnras,
  486, 1907, \dodoi{10.1093/mnras/stz844}

\bibitem[{{Jia} \& {Li}(2015)}]{2015ApJ...814...74J}
{Jia}, K., \& {Li}, X.-D. 2015, \apj, 814, 74,
  \dodoi{10.1088/0004-637X/814/1/74}

\bibitem[{{Jia} \& {Li}(2016)}]{2016ApJ...830..153J}
---. 2016, \apj, 830, 153, \dodoi{10.3847/0004-637X/830/2/153}

\bibitem[{{Koester}(2010)}]{2010MmSAI..81..921K}
{Koester}, D. 2010, \memsai, 81, 921

\bibitem[{{Koleva} {et~al.}(2009){Koleva}, {Prugniel}, {Bouchard}, \&
  {Wu}}]{2009A&A...501.1269K}
{Koleva}, M., {Prugniel}, P., {Bouchard}, A., \& {Wu}, Y. 2009, \aap, 501,
  1269, \dodoi{10.1051/0004-6361/200811467}

\bibitem[{{Li} {et~al.}(2022){Li}, {Wang}, {Zhao}, {Bai}, {Yuan}, {Zhang}, \&
  {Liu}}]{2022ApJ...938...78L}
{Li}, X., {Wang}, S., {Zhao}, X., {et~al.} 2022, \apj, 938, 78,
  \dodoi{10.3847/1538-4357/ac8f29}

\bibitem[{{Li} {et~al.}(2019){Li}, {Chen}, {Chen}, \&
  {Han}}]{2019ApJ...871..148L}
{Li}, Z., {Chen}, X., {Chen}, H.-L., \& {Han}, Z. 2019, \apj, 871, 148,
  \dodoi{10.3847/1538-4357/aaf9a1}

\bibitem[{{Lindegren} {et~al.}(2021){Lindegren}, {Bastian}, {Biermann},
  {Bombrun}, {de Torres}, {Gerlach}, {Geyer}, {Hern{\'a}ndez}, {Hilger},
  {Hobbs}, {Klioner}, {Lammers}, {McMillan}, {Ramos-Lerate},
  {Steidelm{\"u}ller}, {Stephenson}, \& {van Leeuwen}}]{2021A&A...649A...4L}
{Lindegren}, L., {Bastian}, U., {Biermann}, M., {et~al.} 2021, \aap, 649, A4,
  \dodoi{10.1051/0004-6361/202039653}

\bibitem[{{Lomb}(1976)}]{1976Ap&SS..39..447L}
{Lomb}, N.~R. 1976, \apss, 39, 447, \dodoi{10.1007/BF00648343}

\bibitem[{{Masci} {et~al.}(2019){Masci}, {Laher}, {Rusholme}, {Shupe}, {Groom},
  {Surace}, {Jackson}, {Monkewitz}, {Beck}, {Flynn}, {Terek}, {Landry},
  {Hacopians}, {Desai}, {Howell}, {Brooke}, {Imel}, {Wachter}, {Ye}, {Lin},
  {Cenko}, {Cunningham}, {Rebbapragada}, {Bue}, {Miller}, {Mahabal}, {Bellm},
  {Patterson}, {Juri{\'c}}, {Golkhou}, {Ofek}, {Walters}, {Graham}, {Kasliwal},
  {Dekany}, {Kupfer}, {Burdge}, {Cannella}, {Barlow}, {Van Sistine}, {Giomi},
  {Fremling}, {Blagorodnova}, {Levitan}, {Riddle}, {Smith}, {Helou}, {Prince},
  \& {Kulkarni}}]{2019PASP..131a8003M}
{Masci}, F.~J., {Laher}, R.~R., {Rusholme}, B., {et~al.} 2019, \pasp, 131,
  018003, \dodoi{10.1088/1538-3873/aae8ac}

\bibitem[{{McMahon} {et~al.}(2013){McMahon}, {Banerji}, {Gonzalez}, {Koposov},
  {Bejar}, {Lodieu}, {Rebolo}, \& {VHS Collaboration}}]{2013Msngr.154...35M}
{McMahon}, R.~G., {Banerji}, M., {Gonzalez}, E., {et~al.} 2013, The Messenger,
  154, 35

\bibitem[{{Mu} {et~al.}(2022){Mu}, {Gu}, {Yi}, {Zheng}, {Sou}, {Bai}, {Zhang},
  {Lei}, \& {Li}}]{2022SCPMA..6529711M}
{Mu}, H.-J., {Gu}, W.-M., {Yi}, T., {et~al.} 2022, Science China Physics,
  Mechanics, and Astronomy, 65, 229711, \dodoi{10.1007/s11433-021-1809-8}

\bibitem[{{Nelson} {et~al.}(2004){Nelson}, {Dubeau}, \&
  {MacCannell}}]{2004ApJ...616.1124N}
{Nelson}, L.~A., {Dubeau}, E., \& {MacCannell}, K.~A. 2004, \apj, 616, 1124,
  \dodoi{10.1086/421698}

\bibitem[{{Ochsenbein} {et~al.}(2000){Ochsenbein}, {Bauer}, \&
  {Marcout}}]{2000A&AS..143...23O}
{Ochsenbein}, F., {Bauer}, P., \& {Marcout}, J. 2000, \aaps, 143, 23,
  \dodoi{10.1051/aas:2000169}

\bibitem[{{Paxton} {et~al.}(2011){Paxton}, {Bildsten}, {Dotter}, {Herwig},
  {Lesaffre}, \& {Timmes}}]{2011ApJS..192....3P}
{Paxton}, B., {Bildsten}, L., {Dotter}, A., {et~al.} 2011, \apjs, 192, 3,
  \dodoi{10.1088/0067-0049/192/1/3}

\bibitem[{{Paxton} {et~al.}(2013){Paxton}, {Cantiello}, {Arras}, {Bildsten},
  {Brown}, {Dotter}, {Mankovich}, {Montgomery}, {Stello}, {Timmes}, \&
  {Townsend}}]{2013ApJS..208....4P}
{Paxton}, B., {Cantiello}, M., {Arras}, P., {et~al.} 2013, \apjs, 208, 4,
  \dodoi{10.1088/0067-0049/208/1/4}

\bibitem[{{Paxton} {et~al.}(2015){Paxton}, {Marchant}, {Schwab}, {Bauer},
  {Bildsten}, {Cantiello}, {Dessart}, {Farmer}, {Hu}, {Langer}, {Townsend},
  {Townsley}, \& {Timmes}}]{2015ApJS..220...15P}
{Paxton}, B., {Marchant}, P., {Schwab}, J., {et~al.} 2015, \apjs, 220, 15,
  \dodoi{10.1088/0067-0049/220/1/15}

\bibitem[{{Paxton} {et~al.}(2018){Paxton}, {Schwab}, {Bauer}, {Bildsten},
  {Blinnikov}, {Duffell}, {Farmer}, {Goldberg}, {Marchant}, {Sorokina},
  {Thoul}, {Townsend}, \& {Timmes}}]{2018ApJS..234...34P}
{Paxton}, B., {Schwab}, J., {Bauer}, E.~B., {et~al.} 2018, \apjs, 234, 34,
  \dodoi{10.3847/1538-4365/aaa5a8}

\bibitem[{{Paxton} {et~al.}(2019){Paxton}, {Smolec}, {Schwab}, {Gautschy},
  {Bildsten}, {Cantiello}, {Dotter}, {Farmer}, {Goldberg}, {Jermyn}, {Kanbur},
  {Marchant}, {Thoul}, {Townsend}, {Wolf}, {Zhang}, \&
  {Timmes}}]{2019ApJS..243...10P}
{Paxton}, B., {Smolec}, R., {Schwab}, J., {et~al.} 2019, \apjs, 243, 10,
  \dodoi{10.3847/1538-4365/ab2241}

\bibitem[{{Pelisoli} {et~al.}(2018){Pelisoli}, {Kepler}, \&
  {Koester}}]{2018MNRAS.475.2480P}
{Pelisoli}, I., {Kepler}, S.~O., \& {Koester}, D. 2018, \mnras, 475, 2480,
  \dodoi{10.1093/mnras/sty011}

\bibitem[{{Pelisoli} {et~al.}(2017){Pelisoli}, {Kepler}, {Koester}, \&
  {Romero}}]{2017ASPC..509..447P}
{Pelisoli}, I., {Kepler}, S.~O., {Koester}, D., \& {Romero}, A.~D. 2017, in
  Astronomical Society of the Pacific Conference Series, Vol. 509, 20th
  European White Dwarf Workshop, ed. P.~E. {Tremblay}, B.~{Gaensicke}, \&
  T.~{Marsh}, 447.
\newblock \doarXiv{1610.05550}

\bibitem[{{Price-Whelan} {et~al.}(2017){Price-Whelan}, {Hogg},
  {Foreman-Mackey}, \& {Rix}}]{2017ApJ...837...20P}
{Price-Whelan}, A.~M., {Hogg}, D.~W., {Foreman-Mackey}, D., \& {Rix}, H.-W.
  2017, \apj, 837, 20, \dodoi{10.3847/1538-4357/aa5e50}

\bibitem[{{Pylyser} \& {Savonije}(1988)}]{1988A&A...191...57P}
{Pylyser}, E., \& {Savonije}, G.~J. 1988, \aap, 191, 57

\bibitem[{{Ram{\'\i}rez} {et~al.}(2013){Ram{\'\i}rez}, {Allende Prieto}, \&
  {Lambert}}]{2013ApJ...764...78R}
{Ram{\'\i}rez}, I., {Allende Prieto}, C., \& {Lambert}, D.~L. 2013, \apj, 764,
  78, \dodoi{10.1088/0004-637X/764/1/78}

\bibitem[{{Scargle}(1982)}]{1982ApJ...263..835S}
{Scargle}, J.~D. 1982, \apj, 263, 835, \dodoi{10.1086/160554}

\bibitem[{{Schlafly} \& {Finkbeiner}(2011)}]{2011ApJ..737103}
{Schlafly}, E.~F., \& {Finkbeiner}, D.~P. 2011, \apj, 737, 103,
  \dodoi{10.1088/0004-637X/737/2/103}

\bibitem[{{Schlegel} {et~al.}(1998){Schlegel}, {Finkbeiner}, \&
  {Davis}}]{1998ApJ...500..525S}
{Schlegel}, D.~J., {Finkbeiner}, D.~P., \& {Davis}, M. 1998, \apj, 500, 525,
  \dodoi{10.1086/305772}

\bibitem[{{Skrutskie} {et~al.}(2006){Skrutskie}, {Cutri}, {Stiening},
  {Weinberg}, {Schneider}, {Carpenter}, {Beichman}, {Capps}, {Chester},
  {Elias}, {Huchra}, {Liebert}, {Lonsdale}, {Monet}, {Price}, {Seitzer},
  {Jarrett}, {Kirkpatrick}, {Gizis}, {Howard}, {Evans}, {Fowler}, {Fullmer},
  {Hurt}, {Light}, {Kopan}, {Marsh}, {McCallon}, {Tam}, {Van Dyk}, \&
  {Wheelock}}]{2006AJ....131.1163S}
{Skrutskie}, M.~F., {Cutri}, R.~M., {Stiening}, R., {et~al.} 2006, \aj, 131,
  1163, \dodoi{10.1086/498708}

\bibitem[{{Soethe} \& {Kepler}(2021)}]{2021MNRAS.506.3266S}
{Soethe}, L.~T.~T., \& {Kepler}, S.~O. 2021, \mnras, 506, 3266,
  \dodoi{10.1093/mnras/stab1916}

\bibitem[{{Sun} \& {Arras}(2018)}]{2018ApJ...858...14S}
{Sun}, M., \& {Arras}, P. 2018, \apj, 858, 14, \dodoi{10.3847/1538-4357/aab9a4}

\bibitem[{{Tang} \& {Li}(2021)}]{2021MNRAS.506.3323T}
{Tang}, W., \& {Li}, X.-D. 2021, \mnras, 506, 3323,
  \dodoi{10.1093/mnras/stab1937}

\bibitem[{{Tonry} {et~al.}(2012){Tonry}, {Stubbs}, {Lykke}, {Doherty},
  {Shivvers}, {Burgett}, {Chambers}, {Hodapp}, {Kaiser}, {Kudritzki},
  {Magnier}, {Morgan}, {Price}, \& {Wainscoat}}]{2012ApJ...750...99T}
{Tonry}, J.~L., {Stubbs}, C.~W., {Lykke}, K.~R., {et~al.} 2012, \apj, 750, 99,
  \dodoi{10.1088/0004-637X/750/2/99}

\bibitem[{{Torrealba} {et~al.}(2015){Torrealba}, {Catelan}, {Drake},
  {Djorgovski}, {McNaught}, {Belokurov}, {Koposov}, {Graham}, {Mahabal},
  {Larson}, \& {Christensen}}]{2015MNRAS.446.2251T}
{Torrealba}, G., {Catelan}, M., {Drake}, A.~J., {et~al.} 2015, \mnras, 446,
  2251, \dodoi{10.1093/mnras/stu2274}

\bibitem[{{Warner}(2003)}]{2003cvs..book.....W}
{Warner}, B. 2003, {Cataclysmic Variable Stars},
  \dodoi{10.1017/CBO9780511586491}

\bibitem[{{Wilson} \& {Devinney}(1971)}]{1971ApJ...166..605W}
{Wilson}, R.~E., \& {Devinney}, E.~J. 1971, \apj, 166, 605,
  \dodoi{10.1086/150986}

\bibitem[{{Wu} {et~al.}(2014){Wu}, {Du}, {Luo}, {Zhao}, \&
  {Yuan}}]{2014IAUS..306..340W}
{Wu}, Y., {Du}, B., {Luo}, A., {Zhao}, Y., \& {Yuan}, H. 2014, in Statistical
  Challenges in 21st Century Cosmology, ed. A.~{Heavens}, J.-L. {Starck}, \&
  A.~{Krone-Martins}, Vol. 306, 340--342, \dodoi{10.1017/S1743921314010825}

\bibitem[{{Yuan} {et~al.}(2022){Yuan}, {Wang}, {Bai}, {Wang}, {Dong}, {Wang},
  {Yu}, {Zhao}, {Chu}, {Liu}, \& {Zhang}}]{2022ApJ...940..165Y}
{Yuan}, H., {Wang}, S., {Bai}, Z., {et~al.} 2022, \apj, 940, 165,
  \dodoi{10.3847/1538-4357/ac9c62}

\bibitem[{{Zhang} {et~al.}(2022){Zhang}, {Zheng}, {Gu}, {Sun}, {Yi}, {Shi},
  {Wang}, {Bai}, {Zhang}, {Cui}, {Wang}, {Wu}, {Li}, {Shao}, {Lu}, {Bai}, {Li},
  {Fu}, \& {Liu}}]{2022ApJ...933..193Z}
{Zhang}, Z.-X., {Zheng}, L.-L., {Gu}, W.-M., {et~al.} 2022, \apj, 933, 193,
  \dodoi{10.3847/1538-4357/ac75b6}

\bibitem[{{Zhao} {et~al.}(2012){Zhao}, {Zhao}, {Chu}, {Jing}, \&
  {Deng}}]{2012RAA....12..723Z}
{Zhao}, G., {Zhao}, Y.-H., {Chu}, Y.-Q., {Jing}, Y.-P., \& {Deng}, L.-C. 2012,
  Research in Astronomy and Astrophysics, 12, 723,
  \dodoi{10.1088/1674-4527/12/7/002}

\end{thebibliography}

\software{
ATLAS9 \citep[][]{Castelli2003},
Wilson-Devinney binary star modeling code \citep[WD;][]{1971ApJ...166..605W}, 
The Joker \citep{2017ApJ...837...20P}, 
VizieR \citep{2000A&AS..143...23O}.}

\end{document}